\documentclass[pre,superscriptaddress,preprintnumbers,amsmath,amssymb,floatfix, 12pt,authoryear]{revtex4}
\usepackage{graphics}
\usepackage{amsmath}
\usepackage{graphicx,color}
\usepackage[latin1]{inputenc}
\usepackage[T1]{fontenc}
\usepackage{ae,aecompl}

\newcommand \be  {\begin{equation}}
\newcommand \beq {\begin{equation}}
\newcommand \bea {\begin{eqnarray} \nonumber }
\newcommand \ee  {\end{equation}}
\newcommand \eeq {\end{equation}}
\newcommand \eea {\end{eqnarray}}

\newcommand{\beqa}{\begin{eqnarray}}
\newcommand{\eeqa}{\end{eqnarray}}

\newcommand{\ket}[1]{|{#1}\rangle}
\newcommand{\bra}[1]{\langle{#1}|}
\newcommand{\vidk}{\ket{-}}
\newcommand{\vidb}{\bra{-}}
\newcommand{\brak}[2]{\langle{#1}|{#2}\rangle}

\renewcommand{\H}{\mathcal{H}}

\newcommand{\Tr}{\text{Tr}}
\newcommand{\bol}[1]{{\boldsymbol{#1}}}

\begin{document}
\title{Kramers equation and supersymmetry.}

\author{Julien~Tailleur} 
\email{tailleur@pmmh.espci.fr}

\author{Sorin~T\u{a}nase-Nicola}\email{sorin@amolf.nl}
\altaffiliation[present
address:~]{FOM Institute for Atomic and Molecular Physics, Kruislaan
  407, 1098 SJ Amsterdam, The Netherlands}

\author{Jorge~Kurchan}
\email{jorge@pmmh.espci.fr}

\break

\affiliation{ PMMH UMR 7636 CNRS-ESPCI, 10, Rue Vauquelin, 75231 Paris
  CEDEX 05, France}

\date{\today}

\begin{abstract}
  Hamilton's equations with noise and friction possess a hidden
  supersymmetry, valid for time-independent as well as periodically
  time-dependent systems. It is used to derive topological properties
  of critical points and periodic trajectories in an elementary way.
  From a more practical point of view, the formalism provides new tools
  to study the reaction paths in systems with separated time scales. A
  'reduced current' which contains the relevant part of the phase
  space probability current is introduced, together with strategies
  for its computation.
\end{abstract}
\maketitle

\section{Introduction}
\label{sec:intro}

Morse theory describes the relationship between the critical points
of a smooth real (`Morse') function and the topology of the manifold on
which it is defined~\cite{Milnor63}. It is a
major tool in mathematics, but also finds natural applications in
physics, for instance in the classification of periodic orbits in
classical mechanics.

An elementary and elegant derivation of Morse theory was obtained
years ago \cite{Witten82} through the use of supersymmetric quantum
mechanics (SUSY-QM), in which the potential is related to the Morse
function.  It is based on the fact that the semi-classical lowest
eigenstates of the SUSY-QM Hamiltonian are concentrated on the
saddle-points of the function -- those having $k$ fermions on saddles
with $k$ unstable directions. The supersymmetry operators that map
eigenstates with $k$ fermions into eigenstates with $k \pm 1$ fermions
then induce relations between the corresponding saddles.

Soon after this development, it became clear that the SUSY-QM
Hamiltonian -- or, more precisely, its zero-fermion restriction -- is
related by a change of basis to the (Fokker-Planck) equation
describing the evolution of probability associated with a Langevin
process \cite{Nicolai80a,Nicolai80b,Parisi82,Cecotti83}, the
semi-classical limit now becoming the low temperature limit. In this
basis the lowest eigenstates with $k$ fermions are concentrated,
rather than on saddle points with $k$ unstable directions, on the
unstable manifolds emanating from them. For example, one-fermion
eigenvectors with small eigenvalues are peaked on the gradient paths
joining two minima via a saddle point, and so  represent the
reaction current\cite{Tanase04}. 

A low-temperature Langevin process is a practical method to locate
minima of the potential: the dynamics consist of gradient descents,
and the small noise allows to escape local minima - a form of what is
known as `simulated annealing'. In the language of SUSY-QM, such a process
 corresponds to the evolution in zero-fermion subspace. One is then led to ask
whether a dynamics associated with the one-fermion subspace can give a
`simulated annealing' scheme that will converge to reaction paths,
just as the ordinary one does for metastable states. Indeed, this is
so \cite{Tanase03b,Tanase04}, and has been proposed as a basis for
numerical algorithms.

In practical applications it is sometimes necessary to extend the
 overdamped Langevin treatment to a case where inertia plays a role.
 This is possible because pure Hamilton's equations have themselves a
 supersymmetry, whose consequences have been explored extensively by
 Gozzi, Niemi and their co-workers
 \cite{Gozzi93,Deotto03a,Deotto03b,Niemi95,Niemi96,Niemi99}. There is
 however a problem that makes the formalism less transparent than in
 the Langevin case: in classical mechanics the (Liouville) equations
 for the evolution of probability in phase-space do not contain second
 derivatives: the spectrum of the evolution operator is then
 continuous and the space of wavefunctions awkward. A natural cure to
 this problem, both from the analytical and the practical point of
 view, is the introduction of noise and friction so that we study a
 more general process:
\begin{equation}
  \left\{\begin{aligned} \dot q_i&=p_i \\ \dot p_i&=-\frac{\partial
    V}{\partial q_i}-\gamma p_i + \sqrt{2 \gamma T} \eta_i,
    \label{kradyn} \end{aligned}\right.
\end{equation}
$V(\bol{q})$ is a potential energy, $\eta_i$ are Gaussian white
noises modelling the interaction with a thermal bath at temperature $T$,
while $\gamma$ is the coupling to the bath (physically a friction
coefficient).  The dynamics (\ref{kradyn}) can be expressed as a
probability density evolving according to the Kramers equation
\cite{Risken96}:
\begin{equation} \frac{\partial P({\bf q,p},t)}{\partial t}=
\left[ \sum_{i=1}^N \frac{\partial }{\partial p_i} \left(\gamma T
\frac{\partial }{\partial p_i}+\gamma p_i+\frac{\partial V}{\partial
q_i}\right)- \frac{\partial }{\partial q_i}p_i\right] P({\bf
q,p},t)=-H_K P({\bf q,p},t).
\label{FP1}
\end{equation}
It turns out that one can uncover a hidden supersymmetry associated
with this equation, just as there is SUSY-QM associated with the
Fokker-Planck equation without inertia. As we shall see, this leads to
a non-Hermitian supersymmetric quantum mechanics, whose zero-fermion
restriction is the Kramers equation. The higher fermion-number
subspaces contain, again just as in SUSY-QM, the information on the
fixed points and their stable and unstable manifolds, where stability
is now defined with respect to Hamiltonian dynamics perturbed by
friction.  Perhaps more surprising is the fact that one can generalize
these results to the case in which the Hamiltonian depends periodically
on time -- one then has a supersymmetric structure in the Floquet
representation. This supersymmetry has a series of consequences which
we shall also explore.

The organization of this paper is as follows: in Section II we
construct the extension of the Kramers operator and we discuss the
consequences its supersymmetry has on the organization of the
eigenvectors and eigenvalues. In Section III we study the
low-temperature (`semi-classical') limit using standard path-integral
methods, and show that it is dominated by periodic trajectories -- in
particular fixed points -- of the noiseless dynamics.  In Section IV
we use these results to show how one can rederive Morse theory results
using time-independent Hamiltonians. As an alternative, a WKB
treatment of the time-independent case with conservative forces is
possible (Section V): it does not rely on path integrals and is very
close to the treatment of SUSY-QM, with the only complication of
non-Hermiticity. In section VI we present the supersymmetries
associated with a periodically time-dependent Hamiltonian, and give a
first few applications.

Just as the zero fermion subspace corresponds to Kramers' equation, and
this in turn to a process following Hamilton's equations plus noise
and friction, one may ask to what stochastic processes does the
$k>0$ fermion subspaces correspond. In section VII we devise such
processes, and use them to give a constructive (and quite non-rigorous)
derivation of the low-temperature wavefunctions yielding the Morse
complex. As mentioned above, part of the motivation for this work is
the construction of algorithms to find saddle points and reaction
paths between metastable states, a very important problem in Physical
Chemistry. In Section VIII we show the relation between one-fermion
wavefunctions and reaction currents.  Interestingly enough, the
formalism strongly suggests that rather than studying the currents
themselves, a modified `reduced current' should be used, which
contains only the part of the current that is effective in making
transitions, equilibrium circulations within states being
subtracted. Finally, in the Conclusion we outline the several possible
continuations of this work.

\section{SUSY of Kramers dynamics}

\subsection{The  Hamiltonian case}
\label{sec:the}
The Hamiltonian dynamics
\begin{equation}
     \dot q_i=\frac{\partial \cal{H}}{\partial p_i} \qquad ; \qquad
    \dot p_i=-\frac{\partial \cal{H}}{\partial q_i},
\end{equation}
where $\cal{H}$ is the Hamiltonian function, induces an evolution for
probabilities $P({\bf q},{\bf p})$ in phase-space given by
\begin{equation}
  \frac{\partial P}{\partial t}= -H_{\cal{H}}P,\qquad\text{with}\qquad
H_{\cal{H}}=-\sum_{i=1}^N\left(
\frac{\partial \cal{H}}{\partial q_i}\frac{\partial }{\partial p_i}-
\frac{\partial \cal{H}}{\partial p_i}\frac{\partial }{\partial q_i}\right).
\end{equation}
One can uncover a group of symmetries of $H_{\cal{H}}$ by
extending the space with $4N$ fermion operators
$(a_i,a^\dag_i,b_i,b^\dag_i)$, and writing \cite{Gozzi93,Deotto03a,Deotto03b}:
\begin{equation}
 H^S_{\cal{H}}=-\sum_{i=1}^N\left( \frac{\partial \cal{H}}{\partial
q_i}\frac{\partial }{\partial p_i}- \frac{\partial \cal{H}}{\partial
p_i}\frac{\partial }{\partial q_i}\right) + \frac{\partial^2\cal{H}
}{\partial q_i \partial q_j}b^\dag_ia_j- \frac{\partial^2\cal{H}
}{\partial p_i \partial p_j} a_j^\dag b_i+ \frac{\partial^2\cal{H}
}{\partial p_j \partial q_i} (b^\dag_ib_j-a^\dag_ja_i),
\end{equation} 
which reduces to the original $H_{\cal{H}}$ in the zero-fermion
subspace. $H^S_{\cal{H}}$ has a large group of symmetries, generated
by the operator whose action is to multiply by $\cal{H}$ (and other
constants of motion, if present), and by
\begin{eqnarray}
  K&=&\sum_{i=1}^N a_i b_i \;\;\;\quad \quad ; \;\;\;  
  K^\dag=-\sum_{i=1}^N a^\dag_i b^\dag_i  \;\;\quad \quad; \;\;\;  
  F= \sum_{i=1}^N (a^\dag_i a_i+ b^\dag_i b_i) \nonumber \\
  Q_1&=&-{ i}\sum^N_{i=1} \left(\frac{\partial}{\partial q_i}a_i+
    \frac{\partial}{\partial p_i}b_i \right) \;\;\;\qquad\qquad\quad\;\, ; \;\;\;
  { Q}_2 =[K^\dag,Q_1]_- = -{ i}  \sum^N_{i=1} \left(  
    \frac{\partial}{\partial q_i}b_i^\dag -\frac{\partial}{\partial p_i}a_i^\dag  \right) \nonumber\\
  { Q}_3&=&[  Q_2, {\cal{H}}]_- =
  {- i}  \sum^N_{i=1} 
  \left(\frac{\partial \cal{H}}{\partial q_i}b_i^\dag - \frac{\partial \cal{H}}{\partial p_i}a_i^\dag\right) 
  \;\;\; ; \;\;\;
  Q_4 =[ Q_1,{\cal{H}}]_- ={- i}  \sum^N_{i=1} 
  \left(\frac{\partial \cal{H}}{\partial q_i}a_i+\frac{\partial
      \cal{H}}{\partial p_i}b_i \right). \nonumber \\
 \label{alg}
\end{eqnarray}
The supersymmetric charges are nilpotent:
\begin{equation}
Q_1^2=Q_2^{2}= { Q}_3^2= { Q}_4^{2}=0,
\end{equation}
and
\begin{equation}
H^S_{\cal{H}}=(Q_1+{ Q}_2+{ Q}_3)^2.
\end{equation}
As already mentioned in the Introduction, the question of the spectra
of operators and the underlying Hilbert space is rather tricky,
because all operators have at most first order derivatives. Here we
shall work with the Kramers equation, for which all relevant operators
have discrete spectra, the Hilbert space is tractable, but the
symmetry group is considerably smaller.

\subsection{Extended operator and symmetries.}
\label{SUSYch}
Let us go back to the original Kramers dynamics with inertia and
dissipation (\ref{kradyn}), which we shall write in a more general
form \footnote{for a discussion on the different possibilities to
  connect a classical evolution with a thermal bath see
  \cite{Cepas97}}:
\begin{equation}
  \label{gkradyn}
  \left\{\begin{aligned}
  \dot q_i&=\frac{\partial \cal{H}}{\partial p_i} \\ 
  \dot p_i&=-\frac{\partial \cal{H}}{\partial q_i}-\gamma
  \frac{\partial \cal{H}}{\partial p_i} + \sqrt{2 \gamma T} \eta_i.
  \end{aligned}\right.
\end{equation}
Here $\cal{H}$ is a general Hamiltonian function of $\{{\bf q,p} \}$
coordinates, equation (\ref{kradyn}) corresponds to ${\cal{H}}$ of the
particular form ${\cal{H}}={\bf p}^2/{2}+V({\bf q})$. We
shall hence assume that $\H$ is a smooth function, either defined over
a bounded phase-space $\{{\bf q,p} \}$ or growing fast enough at
infinity. The probability density in the phase space evolves as:
\begin{equation}
  \begin{aligned}
    \frac{\partial P({\bf q,p},t)}{\partial t}=&-H_K
  P({\bf q,p},t). \\
  \text{with}\quad H_K=&-\gamma \sum_{i=1}^N
  \frac{\partial }{\partial p_i} \left(T \frac{\partial }{\partial
      p_i}+\frac{\partial \cal{H}}{\partial p_i}\right)-
  \sum_{i=1}^N\left( \frac{\partial \cal{H}}{\partial
      q_i}\frac{\partial }{\partial p_i}- \frac{\partial \cal{H}}{\partial
      p_i}\frac{\partial }{\partial q_i}\right)
\end{aligned}
\label{Kramers}
\end{equation}
One can easily see that the Gibbs density ($e^{-\frac{\cal H}{T}}$) is
the stationary state of the process, the eigenstate of $H_{K}$ with
zero eigenvalue.

In order to construct a fermionic extension we start from the
observation that the previous symmetry charge $Q_1$ is independent of
the dynamics considered, having only a geometrical meaning (it is
related to the so-called exterior derivative). We then propose
\begin{equation}
Q=Q_1=- i \sum^N_{i=1} \left( \frac{\partial}{\partial q_i}a_i+
\frac{\partial}{\partial p_i}b_i \right),
\end{equation}
as one of the supersymmetric charges. By inspection, one can see that
\begin{equation}
  \label{eqn:kramersFPhamiltonian} \begin{aligned}
  H=&H_K+\sum_{i,j=1}^N\left(\frac{\partial^2 {\cal{H}}}{\partial q_i \partial
  q_j}b^\dag_ia_j+ \gamma \frac{\partial^2 {\cal{H}}}{\partial p_i
  \partial p_j} b^\dag_jb_i - \frac{\partial^2 {\cal{H}}}{\partial p_i
  \partial p_j}a_i^\dag b_j+ \frac{\partial^2 {\cal{H}}}{\partial q_i
  \partial p_j}(\gamma b^\dag_ja_i+b^\dag_ib_j -a^\dag_ja_i)\right)
  \end{aligned}
\end{equation}
has $Q$ as a symmetry.  By analogy with the Hamiltonian case one may
 ask if there is a first order differential operator $\bar Q$
 satisfying: 
\begin{equation} \bar Q^2=Q^2=0, \quad T[Q, \bar Q]_+=T
 (Q+\bar Q)^2=H.
\label{susyrel}
\end{equation}
This is so with:
\begin{equation}
  \begin{aligned} 
\bar Q &=-i\sum_{i=1}^{N}\left[b_i^\dag \left(
    \frac{\partial}{\partial q_i}+ \frac{1}{T}\frac{\partial
    {\cal{H}}}{\partial q_i} \right)- a_i^\dag \left( \frac{\partial}{\partial
    p_i}+\frac{1}{T}\frac{\partial {\cal{H}}}{\partial p_i} \right)+ \gamma
    b_i^\dag \left( \frac{\partial}{\partial p_i} +
    \frac{1}{T}\frac{\partial {\cal{H}}}{\partial p_i} \right)\right].
    \end{aligned}
\end{equation}
Thus, we now have a supersymmetric extension of $H_K$ and the
corresponding charges \footnote{These symmetries have already been
  obtained by Kleinert and co workers\cite{Kleinert97}, for an action
  corresponding to an evolution of the form (\ref{kradyn}) in the
  (Lagrangian) path-integral formalism.}. The large algebra of
symmetries (\ref{alg}) of the classical formalism has become much
smaller; just as in the case of SUSY-QM, only $Q$, $\bar Q$ and $F$
are symmetries.

The system (\ref{gkradyn}) can be written in terms of the  variables $(x_1, ..., x_{2N}) = ({\bol{q}}, {\bol{p}})$
in a more compact way as:
\begin{equation}
  \label{eqn:generalisedkramersdynamic}
  \begin{aligned}
    \dot x_i &= -\Omega_{ij} \frac{\partial \H}{\partial x_j} - D_{i
      j} \left( \frac{\partial \H}{\partial x_j} -\sqrt{\frac{2
          T}{\gamma}} \eta_j\right)\qquad \Omega_{ij} &=
    \tiny{\begin{pmatrix} 0_N & -1_N \\ 1_N & 0_N
      \end{pmatrix}}\qquad D_{i j}=\gamma \tiny{\begin{pmatrix} 0 & 0
        \\ 0 & 1_N \end{pmatrix}} .
  \end{aligned}
\end{equation}
When $D_{ij}=0$, (\ref{eqn:generalisedkramersdynamic}) is the usual
simplectic formulation of Hamilton's equation, the second term of the
r.h.s represents the interaction with the bath. With this notation,
the generalized Fokker-Planck Hamiltonian
(\ref{eqn:kramersFPhamiltonian}) becomes:
\begin{equation}
  \begin{aligned}
    H&=-\frac{\partial}{\partial x_i} \left( D_{ij} \left( T \frac{\partial
    }{\partial x_j} + \frac{\partial \H}{\partial
    x_j}\right)+\Omega_{i j}\frac{\partial \H}{\partial
      x_j}\right) +(D_{i
      j}+\Omega_{i j} ) \frac{\partial^2 \H}{\partial x_k\partial x_j} c_i^\dag c_k,\\
    &= H_{K} + A_{ik} c^\dag_i c_k,
  \end{aligned}
\end{equation}
where $(c_1, ..., c_{2N})= (a_1,...,a_N,b_1,...,b_N)$, and we have defined:
\begin{equation}
  A_{ik}=(D_{i
    j}+\Omega_{i j} ) \frac{\partial^2 \H}{\partial x_k\partial x_j}.
\label{A}
\end{equation}
The charges are:
\begin{equation}
  \left\{\begin{aligned}
    Q&= -i \sum_{i=1}^{2 N} \frac{\partial}{\partial x_i} c_i\\
    \bar Q&= -\frac{i}{T} \sum_{i=1}^{2N} (\Omega_{ij}+D_{ij}) \left(T \frac{\partial}{\partial x_j}
    +\frac{\partial \H}{\partial x_j} \right)
    c_i^\dag = e^{-\beta \H} (-i) \sum_{i=1}^{2N} (\Omega_{ij}+D_{ij})  \frac{\partial}{\partial x_j}
     c_i^\dag\: e^{\beta \H}.
  \end{aligned}\right.
\end{equation}

The notation in this section has been specialized to the case in which
all the phase space velocities derive from a global function
${\cal{H}}$. Let us keep in mind, however, that when this is not the
case, and the dynamics is given by:
\begin{equation}
  \dot x_i = -\Omega_{ij}  \H_j - D_{i
    j} \H_j -\sqrt{\frac{2 T}{\gamma}} \eta_i,
\end{equation}
with $\frac{\partial \H_i}{\partial x_j}=\frac{\partial \H_j}{\partial
  x_i}$ but $\H_j({\bol{x}})$  not globally a gradient, there still
  is a supersymmetry:
\begin{equation}
  \begin{aligned}
\label{HQ'}	
    Q&= -{ i} \sum_{i=1}^{2N} \frac{\partial}{\partial x_i}
    c_i\qquad \bar Q= -\frac{i}{T} \sum_{i=1}^{2N} A_{ij} \left(T
    \frac{\partial}{\partial x_j} + \H_j \right) c_i^\dag  \\
    H&=-\frac{\partial}{\partial x_i} \left( D_{ij} \left( T
    \frac{\partial }{\partial x_j} + \H_j\right)+\Omega_{i j}
    \H_j\right) + A_{i j} c_i^\dag c_k \quad\text{with}\quad
    A_{ik}=(D_{i j}+\Omega_{i j} ) \frac{\partial \H_j}{\partial x_k}.
  \end{aligned}
\end{equation}

\subsection{The spectrum of $H$}
\label{sec:spectrum}
We shall first examine the case of a time-independent Hamiltonian
${\cal{H}}$, including cases like (\ref{HQ'}) where there is no global
potential. Section \ref{sec:floquet} will be devoted to the
time-dependent case while sections \ref{sec:TI} and \ref{sec:WKB} will
be specifically dedicated to the conservative case with a global $\H$.

$H$ is not hermitian, and cannot be taken to a Hermitian form via a
similarity transformation. It acts on functions of the form
\begin{equation}
\ket{{\psi}}=\sum \psi_{i_1,\dots,i_m,j_1,\dots,j_n}({\bf q,p}) 
b^\dag_{i_1}\dots b^\dag_{i_m} 
a^\dag_{j_1}\dots a^\dag_{j_n}\vidk,
\label{astate}
\end{equation} 
where $\vidk$ is the Fermion vacuum.  We have to distinguish right and
left eigenvectors. They are defined as:
\begin{equation}
  H     \ket{{\psi_i^R}}=\lambda_i \ket{{\psi_i^R}}\qquad\text{and}\qquad
  H^\dag\ket{{\psi_i^L}}=\lambda_i^* \ket{{\psi_i^L}}.
\end{equation}
In what follows we shall suppose that $H$ is diagonalizable i.e. one
can find a bi-orthonormal eigenbasis \footnote{Diagonalizability is not
entirely obvious even in the pure Kramers (zero-fermion) case, see
\cite{Helffer03} for a discussion.}:
\begin{equation}
  \label{biorthbasis}
  \brak{{\psi_j^L}}{{\psi_i^R}}=\delta_{\lambda_i,\lambda_j}.
\end{equation}

\begin{figure}[ht]
  \begin{center}
    \includegraphics{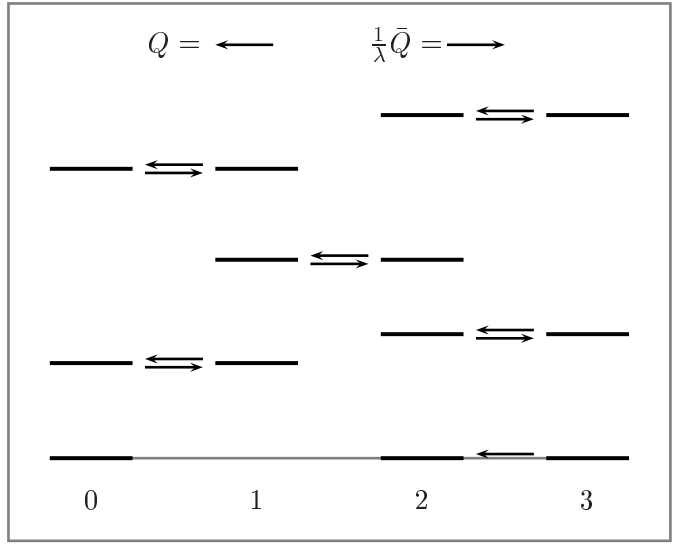}
    \caption{Generic representation of the spectrum of $H$. The
      eigenstates are divided in two kinds: pairs of eigenstates made
      via $Q$ and unpaired eigenstates, which does not belong to any
      pair by $Q$. The latter are only present in the $\lambda=0$
      eigenspace.}
    \label{fig:genericspectrum}
  \end{center}
\end{figure}

The purpose of the two following subsections is to show that the
organization of the spectrum is as in figure
\ref{fig:genericspectrum}.

\subsubsection{Eigenvectors with non-zero eigenvalues}
\label{sec:non0eigenvalues}
Eigenvectors with $\lambda \neq 0$ are not annihilated by both $Q$ and
$\bar Q$. As $Q$ and $\bar Q$ commute with $H$, they map eigenvectors
into eigenvectors with the same eigenvalue. This yields for the
$\lambda \neq 0$ spectrum a structure of degenerate pairs connected by
$Q$ and $\bar Q$. One can indeed construct a basis
$(\ket{{\phi_i^R}},\ket{{\chi_i^R}})$ such that:
\begin{equation} 
 Q \ket{{\phi_i^R}}= \ket{{\chi_i^R}} \qquad\text{and}\qquad 
\bar Q \ket{{\chi_i^R}}=  \frac{\lambda_i}{T} \ket{{\phi_i^R}}.
\label{bases}
\end{equation}

To see this, first notice that a general eigenstate
$\ket{{\psi^R}}$ of eigenvalue $\lambda \neq 0$ is the sum of two
eigenstates $\ket{{\phi^R}}$ and $\ket{{\chi^R}}$ annihilated
by $Q \bar Q$ and $\bar Q Q$ respectively. Indeed, denoting
$\ket{{\chi^R}} \equiv Q \bar Q \ket{{\psi^R}}$ and
$\ket{{\phi^R}} \equiv \bar Q Q \ket{{\psi^R}}$ we have:
\begin{equation}
\ket{{\psi^R}} = \frac{1}{\lambda} H \ket{{\psi^R}} = \frac{T}{\lambda} \left(\ket{{\chi^R}}+\ket{{\phi^R}}\right).
\end{equation}
Using the fact that $Q \bar Q$ and $\bar Q Q$ both commute with $H$,
$\ket{{\chi^R}}$ and $\ket{{\phi^R}} $ are also eigenvectors of
$H$ with eigenvalue $\lambda$, and are annihilated respectively by
$\bar Q Q$ and $Q \bar Q$.  One thus constructs a basis of the whole
eigenspace $\lambda \neq 0$ as the union of bases of eigenvectors
annihilated by $\bar Q Q$ on one hand, and $Q \bar Q$ on the other
hand. We do this as follows: denote $\ket{{\chi_i^R}}$ a basis of
eigenvectors annihilated by $\bar Q Q$, and define
$\ket{{\phi_i^R}}\equiv \frac{T}{\lambda_i} \bar Q \ket{{\chi_i^R}}$.
Because $Q \ket{{\phi_i^R}} = \frac{1}{\lambda_i} H \ket{{\chi_i^R}} =
\ket{{\chi_i^R}}$, the $\ket{{\phi_i^R}}$ so defined are independent,
since one can map them back into an independent set with $Q$.

Furthermore, the $\ket{{\phi_i^R}}$ generate all the eigenvectors
with $\lambda \neq 0$ annihilated by $Q \bar Q$. Indeed, take one such
$\ket{{\psi^R}}$. One clearly has $\ket{{\psi^R}}=
\frac{T}{\lambda} \bar Q Q \ket{{\psi^R}}$.  As $Q
\ket{{\psi^R}}$ is annihilated by $\bar Q Q$, it can be developed
as $Q \ket{{\psi^R}}=\sum_i \alpha_i \ket{{\chi_i^R}}$, and
hence $\ket{{\psi^R}}= \frac{T}{\lambda} \sum_i \alpha_i \bar Q
\ket{{\chi_i^R}} = \frac{1}{\lambda} \sum_i \alpha_i \lambda_i
\ket{{\phi_i^R}}$.

The family $(\ket{{\phi_i^R}},\ket{{\chi_i^R}})$ is thus a
basis of the whole $\lambda_i\neq0$ spectrum, which satisfy the
pairing property (\ref{bases}).  As to the left eigenvectors, the
structure is the same. Constructing a basis with
\begin{equation}
\begin{aligned}	
  \brak{{\phi_i^L}}{{\phi_j^R}}=\delta_{ij} \quad \quad&
  \brak{{\chi_i^L}}{{\chi_j^R}}=\delta_{ij} \\
  \brak{{\phi_i^L}}{{\chi_j^R}}=0 \quad \quad\:\:&
  \brak{{\chi_i^L}}{{\phi_j^R}}=0,
\end{aligned}
\end{equation}
the left eigenvectors are paired according to:
\begin{equation} 
 Q^\dag \ket{{\chi_i^L}}= \ket{{\phi_i^L}} \quad \quad \quad 
\bar Q^\dag \ket{{\phi_i^L}}=  \frac{\lambda_i^*}{T} \ket{{\chi_i^L}}.
\label{basesL}
\end{equation}

\subsubsection{Zero eigenvalues and topology}
\label{sec:0eigenvalues}
The scenario in the $\lambda=0$ eigenspace is a little more complex.
Let us argue that one can in principle build a basis of right
eigenvectors (we drop the index $R$ within this subsection) composed
of (see figure \ref{fig:genericspectrum}):
\begin{equation}
  \label{eqn:basisdef}
  \begin{aligned}
    &\text{ {\em i)} pairs}\;(\ket{{\phi_i^{k+1}}},\ket{{\chi^k_i}})
    \qquad\qquad\qquad\,
    \text{such that}\; \ket{{\chi^k_i}}= Q \ket{{\phi^{k+1}_i}} \neq 0\\
    &\text{ {\em ii)} unpaired eigenstates}\; \ket{{\rho^k_i}}
    \qquad\quad\text{such that}\; Q\ket{{\rho^k_i}}=0\;\text{and}\; \forall
    \ket{{\psi}} \; \ket{{\rho^k_i}} \neq Q \ket{{\psi}},
\end{aligned}
\end{equation}
where $k$ denotes the number of Fermions. Note that such a basis is
matched by the corresponding one for the left eigenvectors.

We shall now construct a basis satisfying (\ref{eqn:basisdef}). Let us
first look at the 0 fermion sector. All the eigenvectors are
annihilated by the $a_i$ and consequently by $Q$. Some of them are the
image by $Q$ of other eigenvectors, and some are not. Let us note
$\ket{\chi_i^0}$ a basis of the former, and $\ket{\phi_i^1}$ the
1-fermion eigenvectors which generate them: $\ket{\chi_i^0}=Q
\ket{\phi_i^1}$. The $\ket{\chi_i^0}$ may not be a basis of the zero
fermion sector and one can complete them with eigenvectors
$\ket{\rho_i^0}$. By definition, the $\ket{\rho_i^0}$ are annihilated
by $Q$ but are not the image of any other eigenvectors by $Q$. At this
point, $\{\ket{\chi_i^0},\,\ket{\rho_i^0}\}$ constitutes a basis of
the 0 fermion sector satisfying (\ref{eqn:basisdef}). One can then
turn to the 1 fermion sector. The part of this sector which is
annihilated by $Q$ can be organized as the $0$ fermion number, that is
there exist a basis $\{\ket{\chi_i^1}, \ket{\rho_i^1}\}$ which generates
this part, and eigenvectors $\ket{\phi_i^2}$ such that
$\ket{\chi_i^1}=Q \ket{\phi_i^2}$. The family $\ket{\phi_i^1}$
introduced above completes the $\{\ket{\chi_i^1},\,\ket{\phi_i^1}\}$ in a
basis of the whole 1 fermion sector. This construction can be followed
for every fermion sector and allows us to construct inductively a
basis $\{\ket{\chi_i^k},\, \ket{\phi_i^k},\,\ket{\rho_i^k}\}$ which
satisfy (\ref{eqn:basisdef}).

In the usual SUSY-QM, only unpaired states exist in the zero
eigenvalue subspace. Indeed, we shall show in Section \ref{sec:WKB}
that this is also the case for a Kramers problem with time-independent
forces that derive from a global potential. Note however, that
states paired by $\bar Q$ with $\lambda=0$ can exist: a system living on a ring
encircling a magnetic flux has no global potential and there is a
pair by $\bar Q$.

The eigenstates $\ket{\rho^k_i}$ are related to the topology of the
phase-space because $Q^\dag$ is the exterior derivative acting on
(left) states with $k$ fermions (the differential $k$-forms). They
span a space whose dimension is the $k^{th}$ Betti number $B_k$, and
is isomorphic to the so-called $k^{th}$ de Rham cohomology group
\cite{Nakahara90} associated with the phase space.
 
The basis of the $\lambda=0$ and $\lambda\neq 0$ eigenspaces form a
global basis: $\{\ket{\psi_i}\}\equiv \{\ket{\chi_i^R},\,
\ket{\phi_i^R},\,\ket{\rho_i^R}\}$.  Given its structure, it is
tempting to denote the states generated by $\{\ket{\rho_i^R}\}$ and
$\{\ket{\phi_i^R}, \ket{\chi_i^R}\}$ as ``unpaired'' and ``paired'',
respectively, even if one can construct states which are not paired by
$Q$ without being generated by the $\ket{\rho_i^k}$.

\section{Low-temperature limit: fixed points and periodic orbits}
\label{sec:trace}

\subsection{Fixed points and periodic orbits}

To study the low-lying eigenvectors in the limit of small $T$, we
shall compute the trace of the evolution operator for different values
of $t$ and use the result to reconstruct the spectrum. A quick method
to do this is to write a path integral and use saddle point
evaluation. This is a standard exercise which we report in Appendix
\ref{app:paths}. When the temperature goes to zero, one finds that
the path integral is dominated by the noiseless periodic trajectories
which satisfy:
\begin{equation}
  \label{eqn:sysperorb}
  \left\{  \begin{aligned}
    \dot p^c_i &= -\frac{\partial \H}{\partial q_i} - \gamma
    \frac{\partial \H}{\partial p_i}\\
    \dot q^c_i &= \frac{\partial \H}{\partial p_i}.\\
  \end{aligned}\right.
\end{equation}
To obtain the next order, one develops the coordinates as a small
perturbation $x_i'=(q_i',p_i')$ around each noiseless orbits
$x^c_i=(q_i^c,p_i^c)$
\begin{equation}
   x_i=x_i^c+\sqrt{T} x_i'.
\end{equation}
The contribution of each orbit ${\bol{x}^c}$ can be seen as the trace
of ${\cal T} e^{-\int_0^t H^{\text{c}}(t') dt'}$, with:
\begin{equation}
  \begin{aligned} H^{\text{c}}&= \frac{\partial}{\partial q'_i} \left(
      p'_j \left.\frac{\partial^2 \H}{\partial p_i \partial
          p_j}\right|_{q^c,p^c} + q'_j\left.\frac{\partial^2 \H}{\partial
          p_i \partial q_j}\right|_{q^c,p^c}\right ) -
    \frac{\partial}{\partial p_i'} \left( \gamma T
      \frac{\partial}{\partial p_i} + \gamma\,p'_j \left.\frac{\partial
          ^2 \H}{\partial p_i \partial p_j}\right|_{q^c,p^c}\right.\\
    &\left. \quad + \gamma\,q'_j \left.\frac{\partial ^2
          \H}{\partial p_i \partial q_j}\right|_{q^c,p^c} +
      p'_j\left.\frac{\partial ^2 \H}{\partial q_i \partial
          p_j}\right|_{q^c,p^c} + q'_j\left.\frac{\partial ^2 \H}{\partial
          q_i \partial
          q_j}\right|_{q^c,p^c}\right)+\left. \frac{\partial^2\cal{H}
      }{\partial q_i \partial 
        q_j}\right|_{q^c,p^c}b^\dag_ia_j\\ &\quad
    +\gamma \left.\frac{\partial^2\cal{H}
      }{\partial p_i \partial p_j}\right|_{q^c,p^c} b_j^\dag b_i   -
    \left.\frac{\partial^2\cal{H} 
      }{\partial p_i \partial p_j}\right|_{q^c,p^c} a_i^\dag b_j +
    \left.\frac{\partial^2\cal{H} }{\partial p_j \partial
        q_i}\right|_{q^c,p^c} (\gamma b_j^\dag a_i + b^\dag_ib_j-a^\dag_ja_i) \\ H^{\text{c}}& =
    -\frac{\partial}{\partial x'_k} \left( D_{kj}
      \frac{\partial}{\partial x'_j} + A^c_{k j}(t) x'_j\right) +
    A^c_{ij}(t) c^\dag_i c_j,
  \end{aligned}
\label{equa}
\end{equation}
where $A^c_{ij}(t)$ is defined as in (\ref{A}), but evaluated along
the classical periodic orbit (which in certain cases will just be a
fixed point):
\begin{equation}
  A^c_{ik}(t) \equiv A_{ik}[x^c(t)]=(D_{i
    j}+\Omega_{i j} ) \left. \frac{\partial \H}{\partial x_k\partial
      x_j}\right|_{x^c}. 
\label{AA}
\end{equation}
Here we have adopted a notation involving a global potential $\H$,
although this is not necessary: if there is no global $\H$ it suffices
to write (\ref{equa}) with $A^c_{k j}$ defined as in (\ref{HQ'}). 

Note that to this order fermionic and bosonic parts are
decoupled. (\ref{equa}) is nothing but the SUSY Hamiltonian
corresponding to a diffusion on a time-dependent harmonic potential
$\H^{\text{c}}=\frac{1}{2} \,\frac{\partial^2 \H}{\partial x_i
\partial x_j} x'_i x'_j$, corresponding to
\begin{equation}
\begin{aligned}
  \dot x'_i &= -A^c_{ij}(t) x'_j + D_i \eta_i,
\end{aligned} 
\label{osc}
\end{equation}
where $\eta_i$ is a Gaussian white noise.

A compact way of expressing the spectral properties of the evolution
operator is via the generating function:
\begin{equation}
T(\lambda,t) \equiv \Tr\left( \lambda^{F} {\cal T} e^{-\int
 H(t') dt'}\right) = \sum_k \lambda^k 
\Tr\left.\left({\cal T} e^{-\int
 H(t') dt'}\right)\right|_{k\;\text{ferm.}},
\label{evol}
\end{equation}
where $ {\cal{T}}$ denotes time order and $F$ is the Fermion number
(\ref{alg}).  To leading order in the temperature $T$, we have to
compute $T(\lambda,t)$ as a sum over the contributions $
T^c(\lambda,t)$ around each periodic orbit: 
\begin{equation}
  T(\lambda,t) \sim \sum_k \lambda^k
  \sum_{\substack{\text{noiseless}\\\text{orbits c}}}
  \Tr\left.\left({\cal T} e^{-\int H^{\text{c}}(t')
        dt'}\right)\right|_{k\;\text{ferm.}}\equiv
  \sum_{\substack{\text{noiseless}\\\text{orbits c}}} T^c(\lambda,t).
  \label{evol1}
\end{equation}
Because boson and fermion degrees of freedom are decoupled to this
order, each term of the sum is a product of a trace over boson and a
trace over fermion degrees of freedom -- both with a (in general
time-dependent) harmonic oscillator.  Again, this is a standard
exercise which we do in detail in Appendix \ref{app:genefunction}.
The result is that for every orbit the boson degrees of freedom
contribute with a factor $|\det(1-U^c(t))|^{-1}$, while the fermion ones
with $\det(1+\lambda U^c(t))$, where $U^c(t)$ is a $2N \times 2N$
matrix defined by:
\begin{equation}
  \dot U^c(t')= -A^c(t')  U^c(t') \quad \quad \quad  U^c(0)=1.
  \label{UU}
\end{equation}

We have then, to leading order in $T$:
\begin{equation}
T(\lambda,t) \underset{T \rightarrow 0}{\longrightarrow} 
\sum_{\substack{\text{noiseless}\\\text{orbits c}}} 
\frac{\det(1+\lambda U^c(t))}{\left|\det (1-U^c(t))\right|}.
\label{generating}
\end{equation}
This formula has two limitations, both reflecting important features
of the phase-space structure:

\begin{itemize}

\item If an eigenvalue of $U^c$ is 1 for some orbits,
(\ref{generating}) diverges. This can be accidental, e.g. a critical
point that is undergoing a second order phase transition, or, more
importantly, a consequence of the fact that the orbit is not isolated
but belongs to a continuous family, possibly as a result of a
symmetry: the problem becomes one of degenerate Morse
theory\footnote{For a time-independent system with non-conservative
forces for example: a single periodic (non-constant) trajectory is in
fact a continuous family of trajectories, each one being the same
orbit with a different starting point.}.

\item The saddle point evaluation is legitimate to the extent that $T
\rightarrow 0$ at fixed $t$.  If we are interested in orbits of period
going to infinity as $T \rightarrow 0$, we have to bear in mind that
the action itself will depend on $T$, and (\ref{generating}) may be
invalid. This problem is not specific to our treatment : the long time
periodic orbits are a usual pitfall in semi-classical quantization.

\end{itemize}

\subsection{Spectrum of low real eigenvalues}
Let us analyze first the contribution to (\ref{generating}) of an
 isolated orbit whose period equals the period $\tau$ of the
 Hamiltonian:
\begin{equation}
  \label{eqn:genelambda}
  T^c(\lambda,n\tau)\sim   \prod_{i=1}^{2N} \frac{ 1+\lambda
    (u_i)^n }{|1-(u_i)^n|}.
\end{equation}
where $(u_1,...,u_{2N})$ are the eigenvalues of $U^c(\tau)$ (for
simplicity, we omit the supra-index $c$ labelling the noiseless
trajectory). One can read in the factor $(1+\lambda u_i)$ the
contribution of the Fermion vacuum whose eigenvalue is $1$ plus
$\lambda$ times the contribution of the one-fermion state, whose
eigenvalue is $u_i$. The bosonic counterpart can be read in
$\frac{1}{|1-u_i|}$: we shall expand $\frac{1}{|1-u_i^n|}$ as a
geometric serie from which we shall recognize the spectrum.

\begin{figure}[ht]
  \includegraphics[totalheight=7cm]{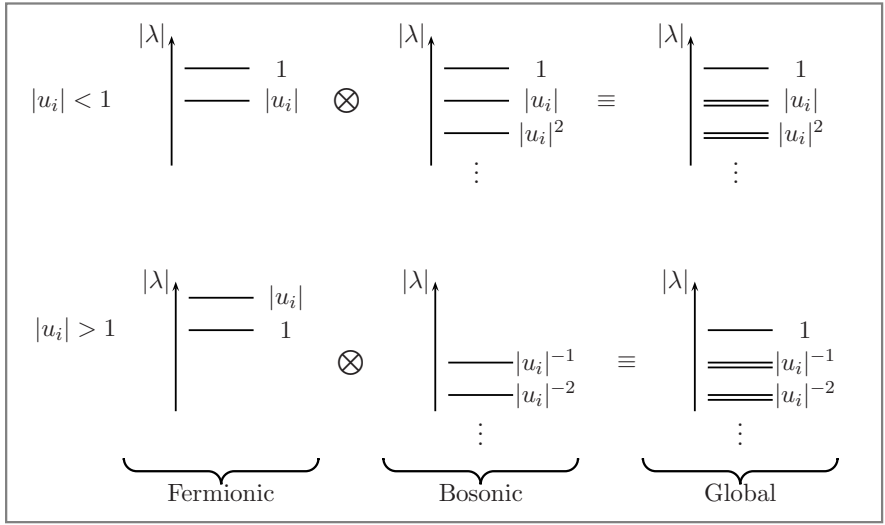}
  \begin{center}
    \caption{Structure of the spectrum for $|u_i|<1$ or $|u_i|>1$ at leading order in $T$. The
      global contribution of the orbit is the tensor product of the
      bosonic  and the fermionic parts. There is a gap
      of order one in  modulus between the most stable state and the first
      excited one when $T$ goes to zero. }
    \label{fig:oscillspectrum}
  \end{center}
\end{figure}

 First consider the case of a real $u_i$. If $|u_i|<1$, then
\begin{equation}
  \frac{1}{|1-u_i^n|}=\frac{1}{1-u_i^n} = \sum_{k=0}^\infty u_i^{n k},
\end{equation}
If $|u_i|>1$, one can rewrite $|1-u_i|$ as $|u_i|(1- \frac{1}{u_i})$.
The same development gives:
\begin{equation}
  \frac{1}{|1-u_i^n|}= \text{sign}(u_i) \sum_{k=1}^\infty u_i^{-kn}  , \quad \forall k \geq 1.
\end{equation}
Next, consider the case of a pair of complex eigenvalues $(u_i,u_i^*)$.
If $|u_i|<1$ we may write:
\begin{equation}
  \frac{1}{|1-u_i^n|} \frac{1}{|1-u_i^{* n}|} = \frac{1}{(1 - u_i^n)}
  \frac{1}{(1 -u_i^{* n})} = \sum_{k=0,k'=0 }  (u_i)^{n k} {(u_i^*)}^{n k'},
\end{equation}
while, if $|u_i|>1$:
\begin{equation}
  \frac{1}{|1-u_i^n|} \frac{1}{|1-u_i^{* n}|} =  \sum_{k=1,k'=1 }  (u_i)^{- n k} {(u_i^*)}^{- n k'}.
\end{equation}
All in all, the contribution of the orbit to the spectrum is the
tensor product of the following sets (see figure \ref{fig:oscillspectrum}):
\begin{itemize}
\item $(1,u_i)\otimes(1,u_i,u_i^2,...)$ for $u_i=u_i^*$, $|u_i|<1$;
\item $(1,u_i)\otimes(\frac{1}{u_i},\frac{1}{u_i^2},...)$ for
  $u_i=u_i^*$, $|u_i|>1$;
\item $(1,u_i)\otimes(1,u_i,u_i^2,...)\otimes
  (1,u_i^*)\otimes(1,u_i^*,{u_i^*}^2,...)$ for $u_i\neq u_i^*$,
  $|u_i|<1$;
\item
  $(1,u_i)\otimes(\frac{1}{u_i},\frac{1}{u_i^2},...)\otimes(1,u_i^*)
  \otimes (\frac{1}{u_i^*},\frac{1}{{u_i^*}^2},...)$ for $u_i\neq
  u_i^*$, $|u_i|>1$.
\end{itemize}
We have supposed here that the number of real eigenvalues $u_i<-1$ is
even. 

Let us now consider the contribution of an orbit of primitive period
$p\tau$.  Clearly, one can start from $p$ different points along the
orbit. For each starting point the preceding discussion holds, and
one gets for the evolution over time $p \tau$ a spectrum as above, but
each level is now $p$-fold degenerate (to this order in $T$). The path
integral tells us that the contribution of this orbit to any trace
over $n$ cycles is zero if $n$ is not a multiple of $p$, and is $p$
times the contribution of a single starting point otherwise. This can
be understood if each multiplet yields, for the evolution over a
single period of the Hamiltonian, eigenvalues corresponding to the $p$
different $p^{th}$ roots of those for $p$ cycles --- because the sum
of the $p$ different roots to the power $n$ is non-zero only if $p$
divides $n$.

Let us stress that breakdown of (\ref{generating}), for example
because one of the $|u_i|=1$, signals the fact that the spectrum is no
longer a superposition of harmonic spectra of frequencies of order
one, and that the gap between ground and first excited state will
go to zero with $T$.  In such cases, the contribution of  the $u_i$
such that $|u_i|\neq 1$ is as described here, and the degrees of
freedom in the directions corresponding to $|u_j|=1$ have to be treated
with other methods (collective coordinates, for example).

The main point of this section is that each orbit is associated with
one and only one eigenstate of the evolution operator with unit
eigenvalue to this order. This eigenstate has $k$ fermions if
$U^c(\tau)$ has $k$ eigenvalues with modulus larger than $1$: that is
to say if its Morse index is $k$.

\section{Time-independent Hamiltonian: A non-hermitian SUSY quantum mechanics and  Morse Theory.}
\label{sec:TI}

In the case of a time-independent Hamiltonian deriving from a global
potential, we recover the Morse inequalities for the stationary points
of the dynamics, by going to the small temperature limit.  The
geometric structures involved now correspond to the manifolds that are
stable and unstable with respect to Hamiltonian dynamics plus
friction, instead of simple gradient descents.

 If ${\cal{H}}$ is time-independent and conservative, in the presence
 of non-zero friction, the only periodic orbits that matter are fixed
 points, as one can see using (\ref{eqn:sysperorb}):
\begin{equation}
\frac{d\,\H}{d\,t}=\sum_i \frac{\partial \H}{\partial q_i} \dot q_i^c +
\frac{\partial \H}{\partial p_i}\dot p_i^c = -
\gamma\,\sum_i \left(\frac{\partial \H}{\partial p_i}\right)^{ 2}.
\end{equation}
Because the energy along an orbit has to be periodic, the only
possibility when $\gamma\neq 0$ is that it is constant and $
\frac{\partial \H}{\partial p_i}=0$.  If the Hamiltonian is of the
form $\H=\frac{1}{2} {\bol p}^2+V(\bol{q})$, this implies that $p_i^c=0$
and $q_i^c=$ constant.  More generally, this implication has to be
verified, but it is true in all but very pathological examples
\footnote{For instance if the points where $\frac{\partial
    \H}{\partial p_i}=0$ are isolated in $p_i$'s direction, then this
  assertion holds. This is for example the case of
  $\H=(p-f(q))^2+V(q)$. In more exotic cases, for instance if the
  phase space is periodic in $p$ and $\H= f(q)\,g(p)+h(q)$ with
  $f(q_0)=f'(q_0)=0$, the solution corresponding to $q=q_0$ and
  $p=h'(q_0)\,t$ is a periodic orbit, which is not localized on a
  critical points.}.

For every fixed point $x^c$, the eigenvalues $u_i^c$ of $U^c(t)$ are
\begin{equation}
u^c_i= e^{-A_i^c t},
\end{equation}
where the $A_i^c$ are the (in general complex) eigenvalues of
$A_{ij}[x_c]$ (Cf. equations (\ref{AA}) and (\ref{UU})). The Morse index of
such a critical point is the number of eigenvalues such that
$|u^c_i|>1$ or equivalently $\text{Re}\,A_i^c<0$. The result of the
preceeding section implies on one hand that all the moduli of the
eigenvalues of $e^{-tH}$ are to this order in $T$ smaller or equal
than one. On the other hand, the number $M_k$ of eigenvalues that
are one within the $k$ fermion subspace coincides with the number of
critical points of index $k$.

For large $t$, we have then:
\begin{equation}
\lim_{t \rightarrow \infty} T(\lambda,t) = \lim_{t \rightarrow \infty}
\sum_k \lambda^k \Tr\left.\left( e^{-
t H  }\right)\right|_{k\;\text{ferm.}}
  = \sum_k \lambda^k M_k,
\label{dino}
\end{equation}
because eigenvalues of $e^{-t H}$ with moduli smaller than one are
exponentially suppressed as in ordinary SUSY-QM.

On the other hand,
\begin{equation}
\begin{aligned}	
T(\lambda,t) &= \sum_{k=0}^{2N} \lambda^k \Tr\left.\left( e^{-
t H }\right)\right|_{k\;\text{ferm.}}\\  &= \sum_{k=0}^{2N} \lambda^k
\Tr\left.\left( e^{-t  H
}\right)\right|_{k\;\text{ferm.}}^{\text{unpaired}} + \sum_{k=0}^{2N}
\lambda^k \Tr\left.\left( e^{- t  H
}\right)\right|_{k\;\text{ferm.}}^{\text{paired}},
\end{aligned}
\end{equation}
where the supraindices denote traces taken over subspaces spanned by
``paired'' and ``unpaired'' states, as defined in section
\ref{sec:0eigenvalues}. We now introduce the partial traces:
\begin{equation}
R_k(t) \equiv \Tr\left.\left( e^{- t H
}\right)\right|_{k\;\text{ferm.}}^{\text{paired states anihilated by } \bar Q}.
\end{equation}
Taking into account the pairing of the spectrum (figure
\ref{fig:genericspectrum}), the fact that the eigenvalues of the
unpaired eigenstates are zero and that the dimension of the space they
generate gives the Betti numbers (see Section \ref{sec:0eigenvalues}),
we get:
\begin{equation}
\begin{aligned}
T(\lambda,t)&= \sum_{k=0}^{2N} \lambda^k B_k + \sum_{k=0}^{2N}
\lambda^k \big(R_k(t) + R_{k-1}(t)\big)\\ \lim_{t\rightarrow
\infty}T(\lambda,t) &=\sum_{k=0}^{2N} \lambda^k B_k + \sum_{k=0}^{2N}
\lambda^k \big(R_k(\infty) + R_{k-1}(\infty)\big),
\end{aligned}
\label{dino1}
\end{equation}
where the $R_k(\infty)$ are integers: the number of paired eigenstates
of $H$ annihilated by $\bar Q$ having eigenvalue zero to leading order
in $T$. Putting together (\ref{dino}) and (\ref{dino1}), we have:
\begin{equation}
M_k= B_k + R_k(\infty) + R_{k-1}(\infty).
\end{equation}
The positivity of the $R_k(\infty)$ (except for $R_{-1}(\infty)=0$)
constitute the strong Morse inequalities:
\begin{equation}
\forall p\quad  \sum_{k=0}^p (-1)^k B_{p-k} \leq \sum_{k=0}^p (-1)^k M_{p-k}
\end{equation}

It is an easy calculation to show for $\H$ of the form
$\frac{1}{2}{\bol {p}^2} + V({\bol x})$ that the index defined here as
the number of eigenvalues with negative real parts tends to the usual
index defined as the number of negative eigenvalues of the potential
$V$ as $\gamma \rightarrow 0$.

Before concluding this section, let us remark that all we have done
here is valid if $\gamma$ is kept finite as $T \rightarrow 0$. If we
wish to consider $\gamma \rightarrow 0$ together with $T=0$, then
orbits that are not fixed points contribute. This is most easily seen
by considering the action in the path integral of Appendix
\ref{app:paths}, which can be written:
\begin{equation}
{\cal{S}}=\frac{1}{2\gamma T} \int dt \; \left(\dot p_i + \gamma
  \,\frac{\partial \H}{\partial p_i}-\frac{\partial \H}{\partial q_i}
  \right)^2 =\frac{1}{2\gamma T} \int dt \; \left(\dot p_i -
  \frac{\partial \H}{\partial q_i} \right)^2 + \frac{\gamma}{2 T} \int
  dt \; {\dot q_i}^2 -\int dt \;\frac{\partial {\cal{H}}}{\partial t},
\label{action}
\end{equation}
restricted to periodic trajectories such that $\dot{q}_i =
\frac{\partial \H}{\partial p_i}$.  If the Hamiltonian is
time-independent, the term $\frac{\gamma}{2T}\int dt \; {\dot q_i}^2$
suppresses periodic orbits that are not fixed points.  However, if we
consider $\gamma \rightarrow 0$, $T \rightarrow 0$, keeping
$\frac{\gamma}{T}$ finite, then non-fixed orbits will contribute, and
we may get richer inequalities.  We shall not follow this strategy
here, although it seems a promising line of research.

\section{A WKB approach.}
\label{sec:WKB}
In this section we treat a  Hamiltonian of the form:
\begin{equation}
  \label{eqn:hamiltonianpdeux}
  {\cal{H}}({\bf q,p})= \sum_i \frac{p_i^2}{2} + V(\bol {q}),
\end{equation}
with a standard WKB treatment that allows to derive Morse theory with
a construction very close to the one in SUSY-QM.

\subsection{Conservative forces}

For Langevin processes (without inertia), the Fokker-Plank operator is
non-Hermitian but can be brought to a Hermitian form by a symmetric
real transformation \cite{Risken96,Tanase04}, provided the forces are
conservative. This is true also for the extension with fermions, and
the transformation yields the Hermitian SUSY-QM. In the Kramers case
$H$ cannot be brought to a Hermitian form, and part of the difficulty
with the proofs comes from there. In this section we shall take a
closer look at a case for which the discussion simplifies
considerably, that of Hamiltonians of the form
(\ref{eqn:hamiltonianpdeux}) for which $H$ reads:
\begin{equation}
  H= \sum_{i=1}^N \left[-\gamma T \frac{\partial^2}{\partial p_i^2}  
    -\frac{\partial V}{ \partial q_i} \frac{\partial}{\partial
      p_i} - \gamma - \gamma p_i \frac{\partial}{\partial p_i}  +
    \frac{\partial}{\partial q_i} p_i\right] + \sum_{i,j=1}^N
  \frac{\partial^2 V}{\partial q_i \partial q_j}b^\dag_ia_j+\sum_{i=1}^N
  \left[ \gamma b^\dag_i b_i -  a_i^\dag b_i\right].
\end{equation}
One can easily verify that $H$, $Q$ and $\bar Q$  are then related to their adjoint by
\footnote{ In fact, this symmetry holds generally for systems with
  energy symmetric under the reflection of $\bf p$: $ H({\bf
    q,-p})=\cal H({\bf q,p})$ and conservative forces.}:
\begin{equation}
H^{\dag}=RHR^{-1},\qquad R\bar Q R^{-1}=Q^\dag, \qquad R Q R^{-1} =
\bar Q^\dag,
\label{pseudoherm}
\end{equation} 
where $R$ is a real, Hermitian invertible operator defined as (no
summation inside the square brackets):
\begin{equation}
  \begin{aligned}
    R&=R^\dag=e^{\frac{\cal H}{T}} \,P\, J, \quad\text{where}\quad P
	\ket{{\psi}({\bf q,p})}=\ket{{\psi}({\bf q,-p})}, \\
	J&=\prod_i[1+a^\dag_ib_i+b_i^\dag
	a_i-a^\dag_ia_i-b_i^\dag b_i+\gamma a^\dag_ia_i+
	(\gamma-2)a^\dag_ib^\dag_ia_ib_i]. 
  \end{aligned}
\end{equation}

Equation (\ref{pseudoherm}) implies that
\begin{equation}
\tilde H = {R^{1/2}}^* H {R^{-1/2}}^* = \Big( R^{1/2} H R^{-1/2} \Big)^\dag,
\end{equation}
which means that if $R^{1/2}$ were real, then $\tilde H$ would be
hermitian. However, this is generically not the case. A weaker result
that will allow us to transpose easily most of the structure of the
SUSY-QM case will be shown below: $R$ is positive-definite -- and hence
$R^{1/2}$ real -- {\em{when restricted to the eigenspace of the lowest
eigenvectors}}.

\subsection{Gaussian development}

In the standard SUSY-QM case, it is useful to work in the basis in
which $H$ is Hermitian, mainly because in that basis the low-lying
eigenvectors peak on saddle-points of the potential.  It would seem
that the analogous thing to do here is to go to an intermediate basis
via $R^{1/2}$. As $R^{1/2}$ is in general non-Hermitian, $H$ is not
Hermitian in this basis, only (complex) symmetric ($H^*=H^\dag$). We
shall instead introduce a different basis $\ket{{\psi^{h R}}} =
e^{\frac{\beta \H}{2}} \ket{{\psi^R}}$ and show that the $\ket{{\psi^{h
R}}}$ whose eigenvalues go to zero with the temperature are
finite-variance Gaussians. Let us compute:
\begin{equation}
  \begin{aligned}
    H'&= e^{\frac{\beta \H}{2}} H  e^{-\frac{\beta \H}{2}}\\
    &= - \gamma T \frac{\partial^2}{\partial p_i^2} -\frac{\gamma}{2} +
    \frac{\gamma}{4 T} p_i^2 -\frac{\partial V}{\partial q_i}
    \frac{\partial}{\partial p_i}+p_i \frac{\partial }{\partial q_i}+
    \frac{\partial}{\partial q_i}  p_i + 
    \frac{\partial^2 V}{\partial q_i \partial q_j}b^\dag_ia_j+
    \gamma b^\dag_i b_i - 
    a_i^\dag b_i.
  \end{aligned}
\end{equation}
Although $H'$ is not Hermitian, it will be more tractable than the
original one.  For each saddle point we propose a WKB form for the
lowest eigenvectors: 
\begin{equation}
  \begin{aligned} 
    \ket{{\psi^{h R}}}&=\ket{{\psi_b^{h R}(\bol q,\bol
        p)}}\otimes\ket{{\psi^{h R}_f}}\\ &= e^{
      -\frac{1}{2T}[B^c_{q_i q_j} (q_i-q^c_i) (q_j-q^c_j)+{B^c_{p_i
          q_j}} p_i (q_j-q^c_j)+{B^c_{p_i p_j}} p_i p_j]} \otimes
    \ket{{\psi_f^{h R}}}.
  \end{aligned} 
  \label{BB}
\end{equation}
In appendix \ref{app:gaussiandev} we show that the Gaussian so defined
has finite variance. This is because the matrix ${\bol{B}}$ in
(\ref{BB}) is not singular, {\em as it is for unstable saddle points
in the original basis} (reflecting the fact that in this basis the
lowest eigenfunctions are {\em not} concentrated on saddles, see
Section \ref{paths}).  We conclude that eigenvectors having eigenvalue
zero (to leading order in $T$) are in this basis, just as in the
ordinary SUSY-QM case, Gaussians peaked on saddle points: those with
$k$ fermions on saddles of index $k$.  In this case, however, right
and left eigenvectors do not coincide.

\subsection{Zero-eigenvalue subspace}

In ordinary SUSY-QM, eigenvectors with exactly zero eigenvalue are
annihilated by  both $Q$ and $\bar Q$. This is evident in the basis in
which $H$ is Hermitian and $Q$ the Hermitian conjugate of $\bar Q$. In
the present case, the proof is slightly more complicated.  We start by
writing:
\begin{equation} 
RH = Q^\dag R Q + {\bar Q}^\dag R {\bar Q}.
\end{equation} 
Clearly, the vectors $\ket{{\psi}}$ annihilated
by $H$ and by $RH$ are the same, and they must satisfy:
\begin{equation}
\label{eqn:jorge}
\langle \psi | R H | \psi \rangle =\langle \psi|Q^\dag R Q| \psi \rangle + \langle \psi|{\bar Q}^\dag R  {\bar Q}| \psi \rangle  =
\langle Q \psi| R | Q \psi \rangle +  \langle {\bar Q} \psi | R
|{\bar Q} \psi \rangle=0.
\end{equation}
If $R$ were positive definite, then this would immediately imply that
both $ Q| \psi \rangle$ and ${\bar Q}| \psi \rangle$ are zero. In
fact, as mentioned above, $R$ is {\em not} positive-definite, but one
can show that it is so when restricted to the subspace of eigenvectors
whose eigenvalues vanish to leading order in $T$ (a space which
obviously contains the eigenstates whose eigenvalues vanish
exactly). To show this, we develop $|\psi \rangle$ as in the previous
subsection:
\begin{equation}
  \ket{{\phi}}=\sum_c \alpha^c e^{-\frac{\beta \H}{2}} \ket{ g^c} \otimes\ket{{f^c}},
\end{equation}
where the sum runs over the critical points `$c$' of $V(q)$. $\ket{ g^c }$
is the normalized Gaussian centered on the  critical point while
$\ket {{f^c} }$ is the corresponding (normalized) fermionic part. Let
us compute $\langle\phi|R|\phi\rangle$:
\begin{equation}
  \begin{aligned}
    \langle\phi|R|\phi\rangle&=\sum_{c,c'} \alpha^c \alpha^{c'*}
    \langle e^{-\frac{\beta \H}{2}} g^c\otimes f^c| e^{\beta \H} P J |
    e^{-\frac{\beta \H}{2}} g^{c'} \otimes f^{c'} \rangle &=
    \sum_{i,j} \alpha^c \alpha^{c'*} \langle g^c | P | g^{c'} \rangle
    \langle f^c | J | f^{c'} \rangle.
  \end{aligned}
\end{equation}
Two Gaussians centered on two different critical points do not overlap
in the small temperature limit:
\begin{equation}
  \langle g^c | P | g^{c'} \rangle=\delta_{c,c'} C^c\quad\text{with}\quad C^c>0,
\end{equation}
and the scalar product reduces to:
\begin{equation}
    \langle\phi|R|\phi\rangle=\sum_c |\alpha^c|^2 C^c
    \langle f^c | J | f^c \rangle.
\end{equation}
It remains to show that the fermion contribution is positive. We can
assume that we have diagonalized the matrix of second derivatives of
the potential at each saddle point, and for each direction $i$ we have
(see Appendix (\ref{app:gaussiandev})) that either $V_{ii}>0$, and then
\begin{equation}
\langle f^c_i | J | f^c_i \rangle=1,
\end{equation}
or $V_{ii}<0$ and
\begin{equation}
  \langle
  f^c_i | J | f^c_i \rangle= 2 \big[\gamma^2 (\gamma+\sqrt{\gamma^2-4V_{ii}})-2
    V_{ii} (2 \gamma + \sqrt{\gamma^2-4 V_{ii}})\big]>0.
\end{equation}
 We have hence shown that in all cases within this subspace either
\begin{equation}
  \langle\phi|R|\phi\rangle>0 \qquad \text{or}\qquad |\phi\rangle=0.
\end{equation}
which, when applied to (\ref{eqn:jorge}) implies that eigenvectors with
exactly zero eigenvalue are unpaired, just as in SUSY-QM.

Paying the price of a loss of generality, this WKB approach shows in a
more intuitive way the organization of the spectrum below the gap. The
eigenstates can be seen -- in an intermediate basis -- as Gaussians
centered on critical points. This Gaussian development enables us to
show that there is no pairing in the zero eigenvalue eigenspace.

\section{Time dependent SUSY}
\label{sec:floquet}
In this section we study the Kramers problem with a periodic
time-dependent Hamiltonian ${\cal H}({\bol{q,p}},t)$ of period
$\tau$.  We first introduce the Floquet formalism and extend the
supersymmetry of the time-independent case to the time-dependent
one. We then briefly show how this formalism allows us to rederive the
Lefschetz formula and even prove the strong Morse inequalities when
the friction is strong enough to prevent the proliferation of orbits
of long periods.

\subsection{Generalized operators}
The Fokker-Planck equation associated with the Kramers system can be written:
\begin{equation}
  \label{eqn:hksdynamics}
  \frac{\partial}{\partial t} \ket{{\psi(\bol q, \bol p;
      t)}}=-H_{K}(\bol q, \bol p; t)\,\ket{{\psi(\bol q, \bol p; t)}}.
\end{equation}
Floquet theory is based on proposing  solutions  through the
ansatz:
\begin{equation}
  \label{eqn:flocquetansatz}	
  \ket{{\psi(\bol q, \bol p; t)}}=\ket{{u(\bol q, \bol p; t)}}\,e^{-\lambda\,t},
\end{equation}
where $\ket{{u(\bol q, \bol p; t)}}$ is periodic of period $\tau$ and
the imaginary part of $\lambda$ can be chosen in the first `Brillouin
zone' $[-\frac{\pi}{\tau},\frac{\pi}{\tau}]$. Equation
(\ref{eqn:hksdynamics}) then becomes:
\begin{equation}
  \label{eqn:hksf}
  \big(H_{K}+\frac{\partial}{\partial t}\big)\,\ket{{u(\bol q, \bol p;
      t)}}=\lambda\, \ket{{u(\bol q, \bol p; t)}}.
\end{equation}

An alternative way to introduce the Floquet representation
\cite{Hanggi91} is to start from the stochastic equation. Introducing
a variable $\theta$ which grows linearly in time, we can write the
Langevin equation (\ref{gkradyn}) as a system evolving with the
time-independent Hamiltonian ${\cal H}({\bol{q,p}},\theta)$:
\begin{equation}
  \label{simplek}
  \left\{\begin{aligned}
  \dot q_i&=\frac{\partial \cal{H}}{\partial p_i} \\
  \dot p_i&=-\frac{\partial \cal{H}}{\partial q_i}-
  \gamma \frac{\partial \cal{H}}{\partial p_i} + \sqrt{2 \gamma T} \eta_i \\
  \dot \theta &= 1,
  \end{aligned}\right.
\end{equation}
with
\begin{equation}
\theta(0)=0.
\label{iit}
\end{equation}
The equation (\ref{eqn:hksf}) leads us to a Kramers operator in the
space $({\bol{q,p}},\theta)$:
\begin{equation}
  \label{Kramerss}
  \mathbb{H}_K=-\gamma \sum_{i=1}^N  \frac{\partial }{\partial p_i} 
  \left(T \frac{\partial }{\partial p_i}-\frac{\partial \cal{H}}{\partial p_i}\right)+
  \sum_{i=1}^N\left(
  \frac{\partial \cal{H}}{\partial q_i}\frac{\partial }{\partial p_i}-
  \frac{\partial \cal{H}}{\partial p_i}\frac{\partial }{\partial q_i}\right)+
  \frac{\partial}{\partial \theta}.
\end{equation}
To construct the supersymmetry of this operator, we first
generalize the operator $Q$ by introducing new fermion operators $a_\theta$ and $a_\theta^\dag$:
\begin{equation}
  \mathbb{Q}\equiv Q-i \,a_\theta \frac{\partial}{\partial \theta}=-
  i \sum^N_{i=1} \left( \frac{\partial}{\partial q_i}a_i+
  \frac{\partial}{\partial p_i}b_i + \frac{\partial}{\partial \theta}
  a_\theta\right).
\end{equation}
We can then extend $\mathbb{H}_K$ by adding  fermion
creation and annihilation operators:
\begin{equation}
  \begin{aligned}
    \label{eqn:mathbbH}
    {\mathbb H}&\equiv \mathbb{H}_K + \frac{\partial^2
      {\cal{H}}}{\partial q_i \partial q_j}b^\dag_ia_j+ \gamma
    \frac{\partial^2 {\cal{H}}}{\partial p_i \partial p_j} b^\dag_ib_j
    -\frac{\partial^2 {\cal{H}}}{\partial p_i \partial p_j}a_i^\dag
    b_j+ \frac{\partial^2 {\cal{H}}}{\partial q_i \partial p_j}(\gamma
    b^\dag_ja_i+b^\dag_ib_j
    -a^\dag_ja_i)\\&\qquad\quad+\left(\frac{\partial^2
        \cal{H}}{\partial q_i\partial \theta}+ \gamma \frac{\partial^2
        \cal{H}}{\partial p_i \partial \theta}\right)b_i^\dag a_\theta
    - \frac{\partial^2 \cal{H}}{\partial p_i \partial\theta}a_i^\dag
    a_\theta.
  \end{aligned}
\end{equation}
$\mathbb H$ commutes with $\mathbb Q$, and we can also construct the
other generator of supersymmetry:
\begin{equation}
  \begin{aligned}	
    \bar {\mathbb Q}=-{i}\left[b_i^\dag
      (\frac{\partial}{\partial q_i} + \gamma \frac{\partial}{\partial
      p_i} +\frac{1}{T}\frac{\partial {\cal{H}}}{\partial q_i}+\gamma
      \frac{1}{T} \frac{\partial {\cal{H}}}{\partial p_i})- a_i^\dag(
      \frac{\partial}{\partial p_i}+ \frac{1}{T} \frac{\partial
      {\cal{H}}}{\partial p_i})-a_\theta^\dag \right],
  \end{aligned}
\end{equation}
satisfying:
\begin{equation}
  \bar {\mathbb Q}^2={\mathbb Q}^2=0, \quad T[{\mathbb Q}, \bar
    {\mathbb Q}]_+=T ({\mathbb Q}+\bar{\mathbb Q})^2=\mathbb H.
  \label{susyrels}
\end{equation}
Let us also note that $a_\theta$ is a symmetry: 
\begin{equation}
[{\mathbb H}, a_\theta]=0,
\end{equation}
implying that if $\ket{{\psi^R}}$ is a right eigenvector of ${\mathbb
  H}$, then $a_\theta\ket{{\psi^R}}$ is a degenerate one, or zero.  In the
following sections, we will use the supersymmetry to study the
eigenvectors and use it to derive relations between the number of
periodic trajectories.

\subsection{Structure of the spectrum -- Quadruplets}

\begin{figure}[ht]
  \includegraphics[totalheight=8cm]{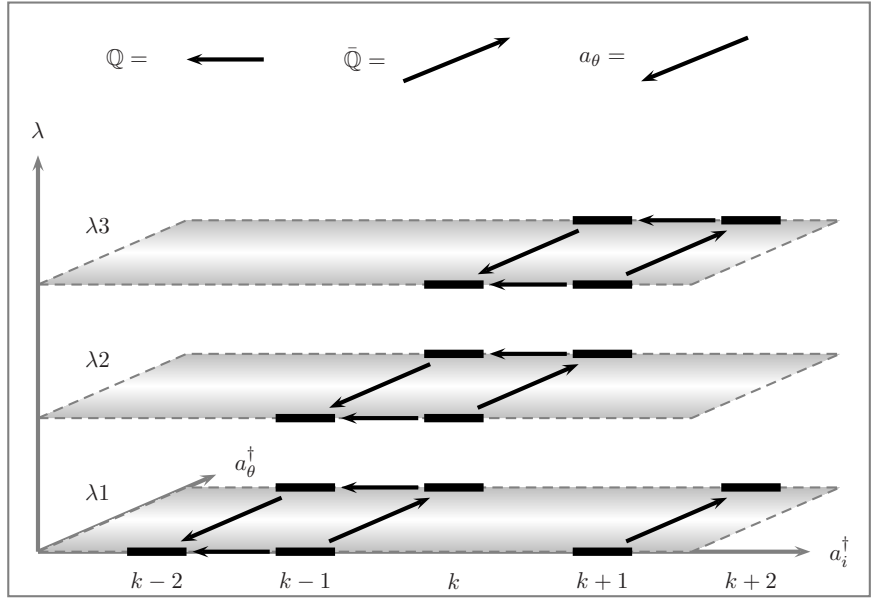} \begin{center}
    \caption{Structure of the spectrum of $\mathbb H$. The vertical
      axis represents the different complex eigenvalues.  The
      horizontal axes represent the number of fermions of type
      $a_\theta^\dag$ and the total number of fermions respectively.
      One can find a basis of the eigenvectors annihilated by
      $a_\theta$ which is organised as in the time independant case.
      It can be completed by a degenerate linearly independent family
      to build a basis of the whole space.}
  \label{fig:quadruplets} \end{center}
\end{figure}

We shall prove here that the right eigenvectors are in fact duplicated in
the following way: one can build a basis
$\{\ket{\psi^0}\}\equiv\{\ket{\phi_i^0},
\ket{\chi_i^0},\ket{\rho_i^0}\}$ of the space annihilated by
$a_\theta$ which satisfy (\ref{eqn:basisdef}) and complete it by a
degenerate free family
$\{\ket{\psi^\theta}\}\equiv\{\ket{\phi_i^\theta},
\ket{\chi_i^\theta},\ket{\rho_i^\theta}\}$, having non zero component
along $a_\theta^\dag$. This basis satisfies (see figure
\ref{fig:quadruplets}):
\begin{equation}
  \label{eqn:quadruplets}
  \begin{aligned}
    a_\theta \ket{\chi_i^0} =0\:\:\:\:\;\qquad& a_\theta \ket{\phi_i^0} =
    0\qquad \quad \:\: a_\theta \ket{\rho_i^0}=0 \\
    a_\theta \ket{\chi_i^\theta} \sim \ket{\chi_i^0} \qquad&
    a_\theta \ket{\phi_i^\theta} \sim \ket{\phi_i^0} \qquad\;
   a_\theta
  \ket{\rho_i^\theta} \sim \ket{\rho_i^0} \\
  {\mathbb Q} \ket{\phi_i^0} = \ket{\chi_i^0}\qquad\,& {\mathbb Q}
  \ket{\rho_i^{0}} = 0\qquad\quad \forall
  \ket{\psi}\; Q \ket{\psi} \neq \ket{\rho_i^{0}}\\
  \mathbb{\bar Q} \ket{\phi_i^0} = \ket{\phi_i^\theta} \,\qquad &
  \mathbb{ Q} \ket{\phi_i^\theta} = \ket{\chi_i^\theta}\qquad \;\mathbb
{\bar Q} \ket{\rho_i^0} = \ket{\rho_i^\theta}
  \end{aligned}
\end{equation}

To construct a basis as (\ref{eqn:quadruplets}), one first remarks
that the procedure followed in section \ref{sec:0eigenvalues} is still
valid in the subspace annihilated by $a_\theta$. One can thus
construct a basis $\{\ket{\rho_i^0}, \ket{\phi_i^0}, \ket{
\chi_i^0}\}$ satisfying (\ref{eqn:basisdef}). Next, let us define
$\{\ket{\psi^\theta}\}$ as:
\begin{equation}
  \begin{aligned}
    \ket{{\phi_i^\theta}}&\equiv \mathbb{\bar Q} \ket{\phi_i^0}= \bar
    Q \ket{{\phi_i^0}} + \frac{i}{T} a_\theta^\dag
    \ket{{\phi_i^0}}\\
    \ket{{\chi_i^\theta}}&\equiv \mathbb{Q} \ket{\phi_i^\theta}= (Q
    \bar Q + \frac{1}{T} \frac{\partial}{\partial \theta} ) \ket{{\phi_i^0}} - \frac{i}{T}
    a_\theta^\dag  \ket{{\chi_i^0}}\\
    \ket{\rho_i^\theta}&\equiv \mathbb{\bar Q} \ket{\rho_i^0}= \bar Q \ket{{\rho_i^0}} +\frac{i}{T} a_\theta^\dag
    \ket{{\rho_i^0}}.
  \end{aligned}
\end{equation}
$a_\theta$ sends $\{\ket{\psi^\theta}\}$ back to $\{\ket{\psi^0}\}$ (up
to constant factors), which proves that $\{\ket{\psi^\theta}\}$ is an
independent family. The whole family is obviously also independent, it can generate any
vector annihilated by $a_\theta$, thanks to the family
$\{\ket{\psi^0}\}$, in particular it can generate the family
$\{\ket{\psi^0}, \ket{a^\dag \psi^0}\}$ which is a basis of
the whole space.

Let us look at the $\{\ket{\rho_i^0}\}$. They are annihilated by $Q$
but not the image by $Q$ of any other eigenvector. The dimension of
the space generated by such eigenvectors in the $k$ fermion sector is
the $k$th Betti number $B_k$ of the phase space $\{\bol p, \bol q\}$ (see
section \ref{sec:0eigenvalues}). The dimension of the space generated
by $\{\ket{\rho_i^\theta}\}$ in the $k$ fermion sector is then equal
to $B_{k-1}$ (see figure \ref{fig:quadruplets}). 

In the following, we will call ``paired'' states and ``unpaired''
states, the eigenvectors generated by
$\{\ket{\chi_i^{0}},\,\ket{\phi_i^{0}},\,\ket{\chi_i^{\theta}},\,\ket{\phi_i^{\theta}}\}$
and $\{\ket{\rho_i^{0}},\,\ket{\rho_i^{\theta}}\}$, respectively.

As in any Floquet problem, the spectrum is organized in Brillouin
zones.  This can be seen directly as follows. Consider the family of
periodic operators $O_m=e^{2\pi i m\theta /\tau}$ with $m$ integer.
Clearly, $[{\mathbb H},O_m]=2\pi i m O_m$, which implies that if
$\ket{{\psi^R}}$ is an eigenvector with eigenvalue $\lambda$ then
$O_m\ket{{\psi^R}}$ is an eigenvector with eigenvalue $\lambda+ 2\pi i
m$.  This redundancy is eliminated if we consider the operator $e^{-n
\tau {\mathbb{H}}}$ restricted to a particular starting point for
$\theta$:
\begin{equation}
  \Tr \left. \left( \langle \theta=n \tau \left|e^{-n \tau {\mathbb{H}}}
      \right|\theta=0 \rangle \right)\right|^{\text{other}} =
    \Tr \; \left[ {\cal{T}} e^{\int_0^{n\tau} H(t) dt} \right].
\label{trace1}
\end{equation}
where `other' means that the trace is taken over all variables
(including fermions) except $\theta$, and $H(t)$ is:
\begin{equation}
  \begin{aligned}
    \label{timedep}
    { H(t)}&\equiv -\frac{\partial}{\partial p_i}\left( \gamma T
      \frac{\partial}{\partial p_i}+\frac{\partial {\cal{H}}}{
        \partial q_i}+ \gamma\frac{\partial {\cal{H}}}{ \partial p_i}
    \right) +
    \frac{\partial}{\partial q_i} \frac{\partial {\cal{H}}}{ \partial p_i} + 
    \frac{\partial^2 {\cal{H}}}{\partial q_i \partial
      q_j}b^\dag_ia_j+ \gamma \frac{\partial^2 {\cal{H}}}{\partial p_i
      \partial p_j} b^\dag_ib_j \\&- \frac{\partial^2 {\cal{H}}}{\partial
      p_i \partial p_j}a_i^\dag b_j+ \frac{\partial^2
      {\cal{H}}}{\partial q_i \partial p_j}(\gamma
    b^\dag_ja_i+b^\dag_ib_j -a^\dag_ja_i) +\left(\frac{\partial^2
        \cal{H}}{\partial q_i\partial t}+ \gamma \frac{\partial^2
        \cal{H}}{\partial p_i \partial t}\right)b_i^\dag a_\theta -
    \frac{\partial^2 \cal{H}}{\partial p_i \partial t}a_i^\dag
    a_\theta.
 \end{aligned}
\end{equation}

\subsection{Trace of the evolution operator}
Let us start by building a generating function just as in the previous
section.  We fix the starting point at $\theta=0$ and compute the
trace:
\begin{equation}
  T(\lambda,n\tau) \equiv  \sum_k \lambda^k  \Tr 
  \left. \left( \langle \theta=n \tau \left| e^{-n \tau {\mathbb{H}}}\right| \theta=0 \rangle
    \right)\right|_{\text{k ferm}}^{\text{other}}  =
  \sum_k \lambda^k  \Tr \;  \left.\left( {\cal{T}} 
      e^{\int_0^{n\tau} H(t) dt} \right)\right|_{\text{k ferm}},
\label{trace2}
\end{equation}
where `other' again means that the trace is over all the variables
except $\theta$.  In the $k$ fermion sector, it is divided in two
parts:
\begin{equation}
  \begin{aligned}
    T_k&=\Tr\left.\left({\cal T} \,e^{-\int_0^{n\tau} {
	H}(t)\,dt}\right)\right|_{k\;\text{ferm.}} 
    = \Tr\left.\left({\cal T} \,e^{-\int_0^{n\tau} {
	H}(t)\,dt}\right)\right|_{k\;\text{ferm.}}^{\text{unpaired}} +
 \Tr\left.\left({\cal T} \,e^{-\int_0^{n\tau} {
	  H}(t)\,dt}\right)\right|_{k\;\text{ferm.}}^{\text{paired}}.\\
  \end{aligned}
\end{equation}
Let us show that unpaired eigenvectors have eigenvalue 0. Suppose that
$H \ket{\rho_i^0}=\lambda \ket{\rho_i^0}$ with $\lambda$ not an
integer multiple of $i\,\frac{2\pi}{\tau}$, that is, $\lambda \neq 0$
inside the first Brillouin zone. As $Q$ and $a_\theta$ annihilate
$\ket{\rho_i^0}$:
\begin{equation}
\label{eqn:lambdaegal0}
\frac{1}{T} H \ket{\rho_i^0} = \mathbb{Q \bar Q} \ket{\rho_i^0}=\frac{\lambda}{T} \ket{\rho_i^0}.
\end{equation}
Since $\mathbb Q= Q -i a_\theta \frac{\partial}{\partial \theta}$,
(\ref{eqn:lambdaegal0}) can be written
\begin{equation}
\label{eqn:lambdaegal01}
Q \mathbb {\bar Q} \ket{\rho_i^0} = \left(\lambda -
\frac{\partial}{\partial \theta}\right)\ket{\rho_i^0}.
\end{equation}
Because $\lambda$ is not zero inside the first Brillouin zone, ${\cal
O}\equiv\lambda - \frac{\partial}{\partial \theta}$ acting on periodic
functions of $\theta$ is invertible. As $[Q,{\cal O}] =0$, one also
has $[Q,{\cal O}^{-1}]=0$. (\ref{eqn:lambdaegal01}) then reads
\begin{equation}
Q \left(\frac{T}{\lambda} {\cal O}^{-1} \mathbb{\bar Q}
  \ket{\rho_i^0}\right )= \ket{\rho_i^0},
\end{equation}
which contradicts the fact that $\ket{\rho_i^0}$ is
unpaired by $Q$. This shows that the unpaired eigenstates have
eigenvalues zero, and the trace over them simply gives their number:
\begin{equation}
  \Tr\left.\left({\cal T} \,e^{-\int_0^{n\tau} {
          H}(t)\,dt}\right)\right|_{k\;\text{ferm.}}^{\text{unpaired}}=B_k+B_{k-1}
\end{equation}
so that:
\begin{equation}
T(\lambda,n\tau) =  (1+\lambda)
  \sum_{k=0}^{2N} \lambda^k B_k + 
\sum_k \lambda^k \Tr\left.\left({\cal T} \,e^{-\int_0^{n\tau} 
	  H(t)\,dt}\right)\right|_{k\;\text{ferm.}}^{\text{paired}}.
\end{equation}
On the other hand, because of the quartet structure,
\begin{equation}
    \label{eqn:Morsetrace1}
 \Tr\left.\left({\cal T} \,e^{-\int_0^{n\tau} 
	  H(t)\,dt}\right)\right|_{k\;\text{ferm.}}^{\text{paired}}  =
    \Tr\left.\left({\cal T} \,e^{-\int_0^{n\tau} {
	H}(t)\,dt}\right)\right|_{k\;\text{ferm.}}^{\text{paired without }\;
	a_\theta^\dag} + 
    \Tr\left.\left({\cal T} \,e^{-\int_0^{n\tau} 
	H(t)\,dt}\right)\right|_{k\;\text{ferm.}}^{\text{paired with}\;
	a_\theta^\dag},
\end{equation}
leading to (see figure \ref{fig:quadruplets}):
\begin{equation}
  \begin{aligned}
  T(\lambda,n\tau)   &= (1+\lambda) \sum_{k=0}^{2N} \lambda^k B_k +   (1+\lambda)
 \sum_k \lambda^k
    \Tr\left.\left({\cal T} \,e^{-\int_0^{n\tau} 
	H(t)\,dt}\right)\right|_{k\;\text{ferm.}}^{\text{without }\;
	a_\theta}   \\
    T(\lambda,n\tau)&= (1+\lambda) \sum_{k=0}^{2N} \lambda^k B_k +
    (1+\lambda) 
    \sum_{k=0}^{2N}\lambda^k ( R_k (n\tau)+R_{k+1}(n\tau)).
  \end{aligned}
\end{equation}
where again we have denoted $R_k(n\tau)$ the partial trace of
$e^{-{n\tau} \mathbb{H}}$ over $k$-fermion states annihilated by
$a_\theta$ and not by $\mathbb Q$.

We now perform the low temperature `semi-classical' evaluation of the
trace.  Because we only need the trace restricted to states without
$a_\theta$, all the calculation in section III carries through
without modifications, since the last two terms in (\ref{timedep})
that are absent in (\ref{eqn:kramersFPhamiltonian}) vanish in this
subspace. We have then:
\begin{equation}
  \begin{aligned} T(\lambda,n\tau) &= \sum_k \lambda^k \left. \Tr \;
    \left( {\cal{T}} e^{ \int_0^{n\tau} H(t) dt }
    \right)\right|_{\text{k ferm}} \\ &= (1+\lambda) \sum_k \lambda^k
    \left.  \Tr \; \left( {\cal{T}} e^{ \int_0^{n\tau} H(t) dt }
    \right)\right|_{\text{k ferm}}^{\text{ without} \theta} \\ &
    \underset{T \rightarrow 0}{\longrightarrow} (1+\lambda)
    \sum_{\substack{\text{noiseless}\\\text{orbits c}}}
    \frac{\det(1+\lambda U^c(n \tau))}{\left|\det (1-U^c(n
    \tau))\right|}.
\label{trace3} 
\end{aligned}
\end{equation}
so that, finally:
\begin{equation}
  \label{eqn:GeneMorseTD}
\sum_{\substack{\text{noiseless}\\\text{orbits c}}} 
\frac{\det(1+\lambda U^c(n \tau))}{\left|\det (1-U^c(n \tau))\right|}=
	 \sum_{k=0}^{2N} \lambda^k B_k + \sum_{k=0}^{2N} \lambda^k (R_k(n \tau)+R_{k+1}(n \tau)).
\end{equation}

For $\lambda=-1$, this gives the Lefschetz formula. For any $n$:
\begin{equation}
  \sum_{\substack{\text{noiseless}\\\text{orbits }c}}
  \text{sign} \Big(\det\big(1-
  U^c(n \tau)\big)\Big)
	= \sum_{k=0}^{2N} (-1)^k B_k.
\end{equation}

We can also obtain strong Morse inequalities for the case in which the
total number of orbits is finite. This will happen if the friction is
sufficiently strong and the time dependent forces are sufficiently
weak. 

Consider first the case in which there are only orbits of period
$\tau$ (e.g. the case of a dissipative, adiabatic evolution).  We
concentrate on an orbit and its corresponding $U^c(\tau)$ with
eigenvalues $u_{i_1},...,u_{i_{2N}}$. We order them so that
$u_{i_1}...u_{i_r}$ verify $|u_{i_j}|>1$ and
$u_{i_{r+1}},...,u_{i_{2N}}$ verify $|u_{i_j}|<1$, by definition the
Morse index of the trajectory is then equal to $r$.  We compute for
large $n$ the l.h.s. of (\ref{eqn:GeneMorseTD}):
\begin{equation}
\frac{\det(1+\lambda U^c(n \tau))}{\left|\det (1-U^c(n
\tau))\right|}=\sum_k \lambda^k \sum_{j_1,...,j_k}
\frac{u_{j_1}^n...u_{j_k}^n}{\prod_{i=1}^{2N}
|1-u_i^n|}\underset{n\rightarrow \infty}{\sim} \sum_k \lambda^k
\sum_{j_1,...,j_k}\frac{u_{j_1}^n...u_{j_k}^n}{
|u_{i_1}^n...u_{i_r}^n|}=\lambda^r.
\end{equation}
since the only contribution which survives corresponds to
$\{j_1,...,j_k\}=\{i_1,...,i_r\}$ \footnote{ We are assuming again that the number
of real eigenvalues $u_i<-1$ is even, otherwise the contribution is $-\lambda^r$.}.

In the limit $n \rightarrow \infty$, (\ref{eqn:GeneMorseTD}) then becomes:
\begin{equation}
  \sum_{k=0}^{2N} \lambda^k B_k + \sum_k \lambda^k
  (R_{k+1}(\infty)+R_k(\infty))=  
  \sum_k \lambda^k M_k, 
\label{ufff}
\end{equation}
where $M_k$ is the number of orbits with Morse index $k$, and
$R_k(\infty)$ the number of eigenstates of ${\mathbb{H}}$ having zero
eigenvalue (to this order in $T$) annihilated by $a_\theta$ and not by
${\mathbb Q}$.

Although we know of no concrete example, let us now outline how
equation (\ref{ufff}) would be derived for a system with a {\em
finite} number of periodic orbits of several different
periods. Consider again the limit of large $n$, but taken for $n$
prime. The path-integral evaluation of the trace tells us that we have
to sum the contributions of all orbits of period $n \tau$, that is,
the repetitions of $n$ times the orbit of period $\tau$. On the other
hand, as we have seen above, the spectrum of ${\cal T}
e^{-\int_0^{\tau} H(t) dt}$ contains, in the presence of orbits of
prime period $=p\tau$, multiplets proportional (to this order) to a
number times the $p$ different roots of unity, and their contribution
disappears from the trace over $n$ cycles, since $n$ is not a
multiple of $p$. Hence, formula (\ref{ufff}) is still valid, but now
the $R_k(\infty)$ counts only the eigenstates of ${\mathbb{H}}$ having
zero eigenvalue (to this order in $T$) annihilated by $a_\theta$ and
not by $\mathbb Q$ that are not part of a multiplet.

The fact that the $R_k(\infty)$ are positive integers in equation
 (\ref{ufff}) constitute the strong Morse inequalities, valid for
 relatively large $\gamma$ and/or small intensity of the
 time-dependent potential so that there is a finite total number of
 periodic orbits.

\section{Stochastic dynamics in the $k$-fermion sector}
\label{simulation}

In this section we describe a stochastic dynamics that corresponds to
the extension of the Kramers equation to evolution of vectors with
$k$ fermions. We restrict ourselves to the case in which $\H$ is
time-independent, although the generalization is straightforward.  The
purpose of this exercise is twofold: first, as we shall discuss below,
we intend to use this dynamics as a practical method to find reaction
paths and other structures in phase-space, and second, we shall use
the equations obtained to construct explicitly the low-temperature
eigenvectors in the $k$-fermion subspace, thus completing Morse
Theory.  In fact, the derivation is practically identical to the one
for SUSY-QM in \cite{Tanase04}.

We wish to find a stochastic process such that it somehow represents
\begin{equation}
  \frac{\partial }{\partial t} \ket{{\psi_k(\bol x ,t)}} = - H
  \ket{{\psi_k(\bol x, t)}},
\label{coco}
\end{equation}
where $\ket{{\psi_k(\bol x,t)}}$ has $k$ fermions. For large
times, the $\ket{{\psi_k(\bol x,t)}}$ will be a combination of
states whose eigenvalues have small real parts.

For zero fermions the dynamics associated with (\ref{coco}) is clearly
Hamiltonian + noise + friction. Let us discuss the
one-fermion sector in some detail.  A one-fermion wavefunction has the
form $\psi_i(\bol{x}) c^\dag_i \vidk$, in phase-space variables.
Equation (\ref{coco}) reads, for the components $\psi_i(\bol{x})$:
\begin{equation}
\frac{\partial}{\partial t} \psi_i(\bol{x},t) = - H_K \psi_i(\bol{x},t) - A_{ij}  \psi_j(\bol{x},t).
\label{coco1}
\end{equation}
Consider first one particle with a $2N$-component normalized vector
$\bol{u}$ attached to it. The position of the particle evolves as a
Langevin process (\ref{gkradyn}), and the vector $\bol{u}$ as:
\begin{equation}
{\dot{u}}_{i}= -A_{ij}u_{j}+N(\bol{u}) u_{i},
\label{u1} 
\end{equation}
where $N(\bol{u})=\sum_{kl}u_{k}A_{kl} u_{l}$ enforces the constancy
of the norm $\sum_i u_i^2=1$.  The only effect of the vector $\bol{u}$
on the dynamics is that we further impose that each particle has a
creation-annihilation average rate $=-N(\bol{u})$.  From
(\ref{gkradyn}) and (\ref{u1}), we have that the joint distribution
function ${\cal F}(\bol{x},{\bol{u}},t)$ evolves then as:
\begin{equation}
\frac{\partial {\cal F}}{\partial t}=
\left[ -H_{K}- N(\bol{u})+ \frac{\partial}{\partial u_{i}} \left(
\sum_{ij} A_{ij} u_{j} -N(\bol{u})  u_{i} \right) \right] {\cal F}.
\label{uu1}
\end{equation}
One can check, using integration by parts, that
\begin{equation}
  \psi_i({\bol{q}},t) = \int d^N\!\!\bol{u} \; u_i \;  {\cal
    F}(\bol{q},\bol{u},t)
\end{equation}
will evolve according to (\ref{coco1}).

The $k$-fermion generalization is straightforward. The dynamics
 (\ref{coco}) for a vector 
\begin{equation}
 {\bol{\psi}}=\sum_{i_1,\dots,i_k}
 \psi_{i_1,\dots,i_k}(\bol x) c^\dag_{i_1}...c^\dag_{i_k}|- \rangle,
\end{equation}
where the $\psi_{i_1,...,i_k}({\bol {x}})$ are totally antisymmetric,
 reads, in components: 
\begin{equation}
 {\dot{\psi}}_{i_1,\dots,i_k}= - H_K
 {{\psi}}_{i_1,\dots ,i_k} - \sum_\sigma \sum_\alpha \;
 (-1)^{n(\sigma,\alpha)} \; A_{\sigma(i_1),\alpha} \;\;
 \psi_{\sigma(i_2),\dots,\alpha,\dots,\sigma(i_k)},
\label{evolv}
\end{equation}
where $\sigma$ denotes all permutations of $k$ indices, and
 $n(\sigma,\alpha)$ is the sign of the permutation
 $(i_1,i_2,\dots,\alpha,\dots,i_k)\rightarrow
 (\alpha,\sigma(i_1),\sigma(i_2),\dots,\sigma(i_k))$.  Again, the
 particles follow equation (\ref{gkradyn}), while the equations of
 motion for the ${\bol{v}}$ read:
\begin{equation}
{\dot{v}}_{i_1,\dots,i_k}= - 
\sum_\sigma \sum_\alpha\; (-1)^{n(\sigma,\alpha)} \; A_{\sigma(i_1),\alpha}
\; v_{\sigma(i_2),\dots,\alpha,\dots,\sigma(i_k)}+  v_{i_1,\dots,i_k} \; {\cal{N}}(\bol{v}) ,
\label{horror123}
\end{equation}
 with 
\begin{equation}
 {\cal{N}}(\bol{v})= \sum_{i_1,\dots,i_k} v_{i_1,\dots,i_k}
\sum_\sigma \sum_\alpha (-1)^{n(\sigma,\alpha)}\;
A_{\sigma(i_1),\alpha} \;
v_{\sigma(i_2),\dots,\alpha,\dots,\sigma(i_k)}, 
\end{equation}
thus preserving the normalization $\sum_{i_1,\dots,i_k}
v_{i_1,\dots,i_k}^2$. As before, there is cloning with rate
${\cal{N}}(\bol{v})$. It is easy to see that average of
${v}_{i_1,\dots,i_k}$ indeed evolves as the ${{\psi}}_{i_1,\dots
,i_k}({\bol{x}})$.

The equations of motion for $({\bol{v}},{\cal{N}}(\bol{v}))$ give the
expansion rate of a volume element driven by the dynamics.  Given a
point ${\bol x}$, we can write a volume element around it as:
\begin{equation}
\bol{V^k} = \delta \bol{x_1}\wedge \delta \bol{x_2} \wedge  \dots 
\wedge \delta \bol{x_k}\equiv{\cal{M}} \sum_{v_{i_1,\dots,i_k}}v_{i_1,\dots,i_k}
{\hat e}_{i_1} \wedge \dots  \wedge {\hat e}_{i_k},
\label{surface}
\eeq where $\wedge$ is the external (wedge) product, ${\hat e}_i$ are
the basis vectors and the form $v_{i_1,\dots,i_k}$ is normalized.  The
orientation and norm of the volume element evolve along a trajectory
of the particle. Equation (\ref{horror123}) gives the evolution of the
orientation $\bol{ v}$, and ${\cal{N}}={\cal{\dot{M}}}$ gives the
expansion rate.

{\bf{{\large{Low-lying eigenstates of $H$ and geometric structures}}}}

The right eigenvectors $\langle {\bol{x}}|\psi^R_a\rangle$ with 'low'
eigenvalues --- with real part going to zero in the limit $T
\rightarrow 0$ --- are concentrated on the following structures:
\begin{itemize}
\item
For $k=0$: local minima.
\item
For $k=1$: paths originating in saddles of index one, spiralling down (thanks to friction) to local minima.
\item For any $k$: the $k$-dimensional surface generated by all paths
decreasing in energy emanating from a saddle of index $k$.
\end{itemize}
The argument is the same as in the purely dissipative SUSY-QM case
\cite{Tanase04}: we consider that the surface descending from the
saddle of index $k$ is uniformly covered with particles with their
$k$-form $\bol{ v}$ at each point tangential to such a surface.  Both
features are preserved by the evolution, as seen in the previous
subsection. First, as the particles go downhill, the $\bol v$ attached
to them change so as to remain tangential: this is because their
evolution are precisely based on the linearized evolution on the
tangent space.  Secondly, the cloning rate matches exactly the
expansion rate of a small volume advected downhill. Hence, the
distribution of particles descending, and the average value of the
forms $v_{i_1,\dots,i_k}$ attached to them is left invariant by the
$T=0$ dynamics.

\section{ Transition paths. `Reduced current'}
\label{paths}

Part of the motivation for this work has been to use the higher
fermion suspaces to find useful information on phase-space.  In
particular, in the Fokker-Planck SUSY-QM case, the low-lying
eigenvectors of the one fermion subspace yield the transition currents
between states. Here, as we shall see, the formalism itself tells us
that, in order to compute the transition path, the relevant quantity is a
`reduced current', rather than the usual one.  Consider first the
Kramers equation (\ref{Kramers}):
\begin{equation}
  \frac{\partial P({\bf q,p},t)}{\partial t}=-H_K P({\bf
    q,p},t)=-\text{div} \bol J = - \left(\frac{\partial J_{q_i}}{\partial q_i}
    + \frac{\partial J_{p_i}}{\partial p_i} \right),
\label{Kramers1}
\end{equation}
which defines the current:
\begin{equation}
  J_{q_i}= \frac{\partial \cal{H}}{\partial p_i} P({\bf q,p},t) \quad
  \quad \quad \quad J_{p_i}= 
  - \left( \gamma T \frac{\partial }{\partial p_i}+\gamma \frac{\partial
      \cal{H}}{\partial p_i} 
    +\frac{\partial \cal{H}}{\partial q_i}\right)P({\bf q,p},t).
\label{current}
\end{equation}
Inspired by the  Langevin/Fokker-Planck case, we apply the operator $\bar Q$
 to a distribution to obtain a 
current. We get:
\begin{equation}
(-i)T\; {\bar{Q}} P({\bf q,p},t) = \ket{{\psi^R}} \equiv
J^{\text{red}}_{q_i} a^\dag_i \vidk+ J^{\text{red}}_{p_i}
b^\dag_i\vidk,
\end{equation}
where we have defined the {\em reduced current} as:
\begin{equation}
  J^{\text{red}}_{q_i}\equiv J_{q_i} +T \frac{\partial P({\bf
  q,p})}{\partial p_i} \quad \quad \quad \quad J^{\text{red}}_{p_i}= J_{p_i}- T
  \frac{\partial P({\bf q,p})}{\partial q_i}.
\end{equation}
The reduced current has the following good properties:
\begin{itemize}
\item
It differs from the current in a term without divergence, hence the
fluxes over closed surfaces coincide.
\item
It is zero in equilibrium. In a case with metastable states, it is small everywhere, while the true 
current is large within the states.
\end{itemize}
It is easy to prove that $\bar Q$ is the only first-order differential
operator that gives a `current' field with the same fluxes as the
standard current while being zero when applied to the equilibrium
Gibbs state.

In the Fokker-Planck SUSY-QM case, the supersymmetry allows us to
obtain the usual transition current on the basis of $1$-fermion low
lying states. Here, the supersymmetric formalism itself has suggested
a new definition for the current, and a practical way to determine it
(on the basis of simulating equation (\ref{coco1})). The reduced
current is more relevant than the total one as it is directly related
to passages rather than to phase space orbits within a state.

{\large{\bf An example.}}

The example which we have studied is that of one particle evolving in
the one-dimensional double-well potential $V(q)=(q^2-1)^2$.  We have
carried out simulations using the algorithm presented in the section
\ref{simulation} and compared it with a determination of the current
by direct simulation, see figures \ref{T.03g1cf} and \ref{T.03g.1cf}\footnote{The numerical simulation of the
equation (\ref{kradyn}), especially in the low friction limit, should
be implemented carefully, as a finite integration step will produce
large errors. We have used quasi simplectic integrators (see
\cite{Mannella04} and references therein) which provide reliable
results in the whole range of friction we studied.}.

\begin{figure}[ht]
\centering \includegraphics[totalheight=6cm]{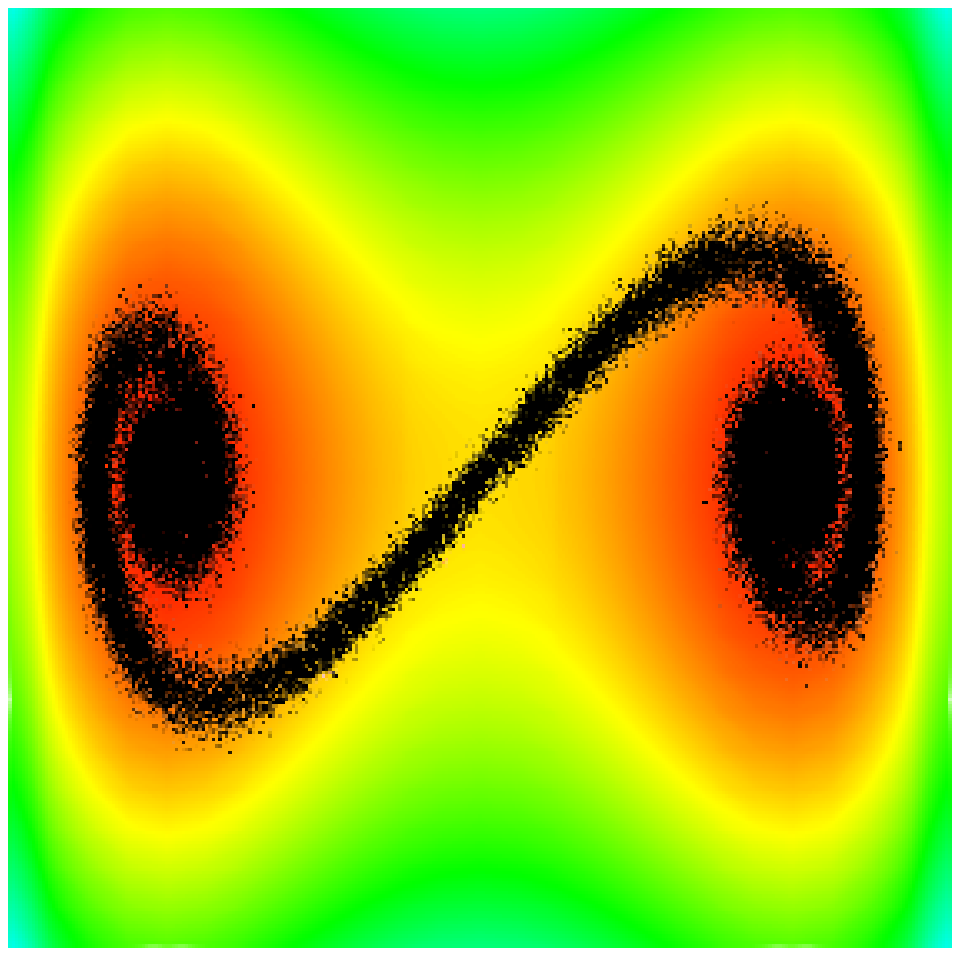}
\caption{Reduced current for the double-well potential of section
\ref{paths}. One can see that the structure is concentrated on two
damped paths, starting in the saddle point, and spiralling down to the
two minima, respectively. The width of the structure is $\sim
\sqrt{\gamma T}$.}
\label{T.03g1cf}
\end{figure}

\begin{figure}[ht]
\centering \includegraphics[totalheight=6cm]{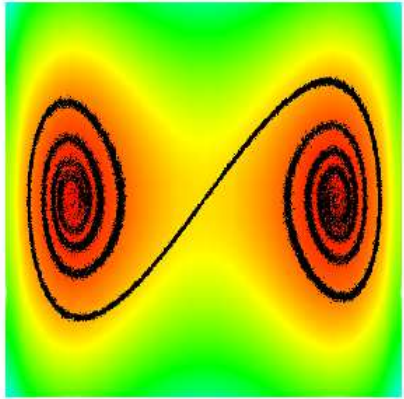}
\caption{Same as figure \ref{T.03g1cf}, with smaller damping}
\label{T.03g.1cf}
\end{figure}

\section{Conclusion and perspectives.}

We have shown how the supersymmetry associated with the Kramers
equation can be used to extract results of Morse theory in an
elementary way. The strategy is very close to the one followed for
the SUSY-quantum mechanics case, although somewhat complicated by the fact
that the operators are non-Hermitian. Compared to the case of pure
Hamiltonian dynamics, we have here a term of friction $\propto
\gamma$, and a term of noise $\propto \sqrt{\gamma T}$ which we can
take to zero in several ways. Thanks to them, the Hilbert space on
which the evolution operator acts is well defined, and the spectra of
the operators are discrete.  The wavefunctions associated with the
eigenvalues with real part close to zero are concentrated on the
structures associated with Morse Theory. 

The same methods can be used for a periodically time-dependent system,
using a supersymmetric structure acting in the Floquet representation
of the problem. The program for this case is however far from
complete, below we mention some possible continuations.

The advantage of this formulation is not only that it is entirely
contained in (physics) undergraduate level, but that it makes a
connection with situations of interest in physics and physical
chemistry.  Although we have used the low-temperature limit as a way
to make the wavefunctions peak on saddles and other structures, other
limiting situations yielding time scale-separation could have been
invoked. For example, in many macroscopic systems, the thermodynamic
states become mutually inaccessible (or almost), and thus play the
same role as minima in the low-temperature case. The formalism
applied here can then be used to build a Morse Theory for states,
their transition currents, and higher objects.  The practical
determination of metastable states and transition paths in this wider
context is an active field of research, especially in complex energy
landscapes relevant for chemical reactions.

We have not dealt here with two elements that are needed in order to
go beyond:
\begin{itemize}

\item Degenerate Morse Theory, non-isolated orbits.

\item Proliferating orbits

\end{itemize}

Degenerate solutions appear in a time-independent system as soon as we
have to consider orbits rather than fixed points, corresponding to the
freedom of choosing the starting point of the orbit. This degeneracy
has already been discussed in the case of SUSY-QM \cite{Witten82}, and
although it makes the treatment more cumbersome, it is a rather
standard exercise in collective coordinates. Two cases when orbits
arise naturally in a time-independent system are when forces do not
derive from a global potential (like a magnetic monopole subjected to
a magnetic field), or when the friction is scaled to zero with the
temperature (e.g. $\gamma/T$ finite).

In systems for which the number of orbits grows exponentially with the
period, a finer method of classifying orbits than the one used in this
paper has to be put in place. This is the subject of Floer Theory,
which we have only marginally touched.

One last development that has been left out here is the case of Hamiltonian
systems with no friction but with thermal noise (for such an infinite
temperature situation, the phase-space has to be finite). Let us just
remark here that in the corresponding low-noise limit, the
supersymmetric formalism provides a method to study the separatrices
and homoclinic orbits.

\appendix

\section{}
\label{app:paths}

The generating function of the evolution operator 
can be written in path integral formalism as \cite{Zinn96}:
\begin{equation}
  \begin{aligned} T(\lambda,t) \equiv \Tr\left( \lambda^{F} {\cal T}
    e^{-\int H(t') dt'}\right) =& \int_{q_0}^{q_0}
    {\cal{D}}[\bol q,\bol p,\bol \eta] \, \delta\left(\dot p_i +
    \gamma\,\frac{\partial \H}{\partial p_i}+\frac{\partial
    \H}{\partial q_i} -\eta_i(t)\right)\\ \delta\left(\dot{q}_i -
    \frac{\partial \H}{\partial p_i}\right)\, &\;
    \exp\left(-\frac{1}{4\,\gamma\,T}\int \sum_i \eta_i^2(t')
    dt'\right) W[q_j,p_j;t'],
    \end{aligned}
\end{equation}
where $W$ is defined by:
\begin{equation}
W[\bol q(t),\bol p(t);t']=\int {\cal{D}}[\bol c,\bol{\bar c}] e^{-\int dt
\,\bar c_i \left(\delta_{ij} (\frac{d}{dt}- \ln \lambda ) +A_{ij}[\bol
p,\bol q]\right) c_i},
\end{equation}
and $(\bar c_j,c_j)$ are Grassmann variables. Using the Fourier
representation of the $\delta$ function \cite{Zinn96}, one gets:
\begin{equation}
  \begin{aligned} 
	T(\lambda,t)= \int_{q_0}^{q_0} {\cal{
D}}[\bol q,\hat{\bol q},\bol p,\hat{\bol p},\bol \eta] e^{\int_0^t dt' \left[
\hat{p}_i \left(\dot p_i + \gamma\,\frac{\partial \H}{\partial
p_i}+\frac{\partial \H}{\partial q_i} -\eta_i(t)\right) -
\frac{1}{4\,\gamma\,T} \sum_i \eta_i^2(t')+\hat{q}_i\left(\dot{q}_i - \frac{\partial \H}{\partial
p_i}\right) \right]} W(\bol q, \bol p;t).
  \end{aligned}
\end{equation}
The question of factor ordering can be dealt with by choosing a
convention in which the integral above represents the Kramers
dynamics. In our case we need not worry about this, as we will only
use the path integral as a bookkeeping device.  The integration over
the noise results in:
\begin{equation}
\label{eqn:toto2}
  T(\lambda,t)=\int_{q_0}^{q_0}
     {\cal D}[\bol q,\hat{\bol q},\bol p,\hat{\bol p}]\, e^{\int dt' \left[
    \hat{p}_i \left(\dot p_i + \gamma\,\frac{\partial \H}{\partial
    p_i}+\frac{\partial \H}{\partial q_i} \right) +
    \hat{p}_i^2\,\gamma\,T+ \hat{q}_i\left(\dot{q}_i -
    \frac{\partial \H}{\partial p_i}\right)\right]} W(\bol q, \bol p;t).
\end{equation}
Integration over $\hat{\bol p}$ and $\hat{\bol q}$ gives a
`Lagrangian' version of the path-integral:
\begin{equation}
  \label{eqn:trace} T(\lambda,t) =\int_{q_0}^{q_0} {\cal D}[\bol q,\bol
p] e^{-\frac{1}{4\,\gamma\,T}\int dt' \left(\dot p_i +
\gamma\,\frac{\partial \H}{\partial p_i}+ \frac{\partial \H}{\partial
q_i} \right)^2 \,}\delta \left(\dot{q}_i - \frac{\partial \H}{\partial
p_i}\right) W(\bol q, \bol p;t).
\end{equation}
The factor $1/\gamma T$ multiplying the action becomes large
when the temperature goes to zero, and the path integral is then dominated by
periodic orbits satisfying:
\begin{equation}
  \label{eqn:Langevinbis}
  \left\{\begin{aligned}
    \dot p_i^c &= - \frac{\partial \H}{\partial q_i} - \gamma
    \frac{\partial \H}{\partial p_i}\\
    \dot q_i^c &= \frac{\partial \H}{\partial p_i}.
  \end{aligned}\right.
\end{equation}
One can linearize equation (\ref{eqn:toto2}) around such an orbit by putting
$x_i=x_i^c+\sqrt{T} x_i'$, $\hat x_i=\frac{1}{\sqrt{T}} \hat x_i'$ and consequently:
\begin{equation}
  \begin{aligned}
    \frac{\partial \H}{\partial x_i} &= \left.\frac{\partial \H}{\partial
      x_i}\right|_{q^c,p^c}+\sqrt{T}\,x'_j\left.\frac{\partial^2 \H}{\partial
      x_i\partial x_j}\right|_{q^c,p^c}.
  \end{aligned}
\end{equation}
One then gets for the contribution of the orbit to this order:
\begin{equation}
  \label{eqn:traceorbite}
  \begin{aligned} 
    &T^c(\lambda,t) = \int {\cal D}[\bol q',\hat{\bol q},\bol p',\hat{\bol p}]
      \,
      e^{-\int dt\, {\hat p}_i \big(\dot p'_i +
      \gamma\,p'_j \left.\frac{\partial ^2 \H}{\partial p_i \partial
      p_j}\right|_{q^c,p^c}+\gamma\,q'_j \left.\frac{\partial ^2
      \H}{\partial p_i \partial q_j}\right|_{q^c,p^c}
      +p'_j\left.\frac{\partial ^2 \H}{\partial q_i \partial
      p_j}\right|_{q^c,p^c}+q'_j\left.\frac{\partial ^2 \H}{\partial
      q_i \partial q_j}\right|_{q^c,p^c}+\gamma {\hat p}_i \big) },\\
&\qquad\qquad e^{-\int dt\, {\hat q }_i \big(\dot q'_i - p'_j
      \left.\frac{\partial^2 \H}{\partial p_i \partial
      p_j}\right|_{q^c,p^c} - q'_j\,\left.\frac{\partial^2
      \H}{\partial p_i \partial q_j}\right|_{q^c,p^c} \big)} W(\bol q^c, \bol p^c;t)
  \end{aligned}
\end{equation}
(the leading contribution to the action vanishes).  Following our
conventions for factor ordering, one recognizes the path
integral representation of the trace of the evolution operator
associated with the time-dependent harmonic Hamiltonian (\ref{equa}).

\section{}
\label{app:genefunction}

We now compute the trace corresponding to the bosonic degrees of
 freedom. Let us assume that the probability distribution is Gaussian:
\begin{equation}
  \label{eqn:defP}
  P(\bol X,t) = \exp\left(-\frac{1}{2}\,B_{ij}(t) \big( X_i-
  X_i^0(t)\big)\big( X_j- X_j^0(t)\big)+C(t)\right),
\end{equation}
where $X_i^0(t)$ has to be determined and $C(t)$ is just a
normalization factor. $P(\bol X,t)$ evolves with (\ref{equa}):
\begin{equation}
  \label{eqn:FPbis}
  \frac{\partial P}{\partial t}=\frac{\partial}{\partial x_k} \left(
  D_{kj} \frac{\partial}{\partial x_j} + A^c_{k j}\, x_j\right) \, P,
\end{equation}
which gives:
\begin{equation}
  \label{eqn:dpadt1}
  \frac{\partial P}{\partial t}=\Big[( X_i- X_i^0)( X_j-
    X_j^0)(BDB-BA^c)_{ij}-( X_i- X_i^0) X_j^0 BA^c_{ij}
    -DB_{kk}+A^c_{kk}\Big] P.
\end{equation}
On the other hand, differentiating directly (\ref{eqn:defP}), we get:
\begin{equation}
  \label{eqn:dpadt2}
  \frac{\partial P}{\partial t}=P\left[-\frac{\dot
      B_{ij}}{2}( X_i- X_i^0)( X_j- X_j^0)+B_{ij}\,\dot
    { X}_i^0( X_j- X_j^0)+\dot C(t)\right].
\end{equation}
Equating (\ref{eqn:dpadt1}) and (\ref{eqn:dpadt2}):
\begin{equation}
  \label{eqn:systemorb}
  \left\{  \begin{aligned}
    -\dot {\bol B}&=2\,(\bol{BDB})-(\bol{BA^c})-(\bol{BA^c})^\dag\\
    \bol B \dot{\bol  X}^0&=-\bol B \bol A^c \,\bol X^0\\
    \dot C(t) &= -\Tr (\bol{DB}-\bol {A^c}).
  \end{aligned}\right.
\end{equation}
The first equation implies for $\bol{B}^{-1}$:
\begin{equation}
\frac{d}{dt}\bol{B}^{-1} = 2 \bol D- \bol A \bol {B}^{-1} -
\bol{B}^{-1} \bol A^\dag, 
\end{equation}
which can be integrated to give:
\begin{equation}
\bol B^{-1}=2\int_0^t \bol U(t) \bol U^{-1}(t') \bol D {\bol
  U^\dag}^{-1}(t') \bol U^\dag(t) dt' + \bol U (t) \sigma_0 \bol
U^\dag (t) .
\end{equation}

Multiplying the second equation by $B^{-1}$
on the left, one gets
\begin{equation}
\label{eqn:timevol1}
\dot{ X}_l^0 = -A^c_{lm} \, X_m^0,
\end{equation} 
which means that $\bol X^0$ follows the noiseless
evolution. Equation (\ref{UU}) tells us that
\begin{equation}
  \label{eqn:def U} X_i^0(t) = U_{ij}^c(t) \,X_j^0(0)
\end{equation}
are solutions of (\ref{eqn:timevol1}).

From the normalization of $P$, we get:
\begin{equation}
  \sqrt{\frac{\det \,\bol B(t)}{(2\pi)^N}}=e^{C(t)},
\end{equation}
which satisfies the third equation of (\ref{eqn:systemorb}). Starting
with $P(\bol X,0)=\delta(\bol X- \bol Y)$, i.e. $\bol X^0(0)=\bol Y$
and $B_{ij}(0)=\underset{\Delta\rightarrow\infty}{\text{lim}} \Delta
\delta_{ij}$, the density is:
\begin{equation}
  \begin{aligned}
     P_{\bol Y}(\bol X,t) &= \sqrt{\frac{\det \,\bol B(t)}{(2\pi)^N}}
     e^{-\frac{B_{ij}}{2}(X_i-U_{ik}^c Y_k)(X_j-U^c_{jl} Y_l)}.
  \end{aligned}
\end{equation}
The trace now reads 
\begin{equation}
\int P_{\bol Y}(\bol Y,t) d\bol Y= \sqrt{\frac{\det \,\bol B}{(2\pi)^N}}
    \sqrt{\frac{(2\pi)^N}{\det\big(1-U^c(t)\big)^\dag\,\det \bol B \,\det
    \big(1-U^c(t)\big)}}, 
\end{equation}
that is:
\begin{equation}
  \label{eqn:boztrace}
\int P_{\bol Y}(\bol Y,t) d\bol Y=\frac{1}{|\det\big(1-U^c(t)\big)|}.
\end{equation}
Therefore,  the bosonic contribution of a classical periodic orbit
is $\frac{1}{|\det(1-U^c)|}$.

The fermionic counterpart to this contribution is obtained by
taking
\begin{equation}
\sum_p \lambda^p \left.  \Tr\left({\cal T} e^{-\int A^c_{ij} c_i^\dag
    c_j}\right)\right|_{\text{p ferm.}}=\sum_p \lambda^p \sum_{i_1,...,i_p} \vidb c_{i_1}...c_{i_p} {\cal T}
    \,e^{-\int A^c_{ij}c^\dag_i c_j}
    c^\dag_{i_p}...c^\dag_{i_1}\vidk,
\end{equation}
where the time-order is defined along the noiseless trajectory. Let us
first calculate the terms $\bra{-}c_{i_1}...c_{i_p} {\cal T} \,e^{-\int A^c_{ij}c^\dag_i
  c_j} c^\dag_{i_p}...c^\dag_{i_1}\vidk$. Using the fact that:
\begin{equation}
  \label{eqn:evolUadag} \left( {\cal T} e^{-\int_0^t A^c_{i j}
c^\dag_i c_j}\right) \,c^\dag_{i_l}\, \left({\cal T} e^{\int_0^t
A^c_{ij} c^\dag_i c_j}\right)=U^c_{i_k i_l}(t)\,c^\dag_{i_k},
\end{equation}
which can easily be seen by differentiating right and left with respect to
$t$, and denoting $O={\cal T} e^{-\int A^c_{i j}\,c^\dag_i\,c_j}$, we
get:
\begin{equation}
  \begin{aligned}
    \vidb c_{i_1}...c_{i_p} O c^\dag_{i_p}...c^\dag_{i_1} \vidk
    &=\vidb c_{i_1},...c_{i_p} O c^\dag_{i_p} O^{-1}\,O c^\dag_{i_{p-1}}
    O^{-1}\,...O c^\dag_{i_1} O^{-1} \vidk\\
    &=\vidb c_{i_1},...c_{i_p} \sum_{j_p} U^c_{j_p i_p}
    c^\dag_{j_p}\sum_{j_{p-1}} U^c_{j_{p-1} i_{p-1}}
    c^\dag_{j_{p-1}}...\sum_{j_1} U^c_{j_1 i_1} c^\dag_{j_1}\vidk\\
    &=\sum_{j_1...j_p} \prod_{k=1}^p U^c_{j_k i_k}
    \bra{-}c_{i_1},...c_{i_p}c^\dag_{j_p},...c^\dag_{j_1}\vidk.
  \end{aligned}
\end{equation}
For a bracket to be non-zero, $j_1,\cdots ,j_p$ must be a permutation of
$i_1,\cdots ,i_p$. The scalar product is then the sign of the permutation,
and one has:
\begin{equation}
  \label{eqn:tracefermsect}
  \vidb c_{i_1},...c_{i_p} O
  c^\dag_{i_p},...c^\dag_{i_1}\vidk ={\det}_p \,U^c_{i_1,...,i_p},
\end{equation}
where $\det_p\,U^c_{i_1,...,i_p}$ is the minor of order $p$ of
$U^c(t)$ associated with the directions $i_1,...,i_p$. We can now
compute the generating function of the fermionic orbit's contribution:
\begin{equation}
\begin{aligned}	
  \label{eqn:defgeneferm}
\sum_p \lambda^p \Tr \left.{\cal T} e^{-\int_0^t
      A^c_{i j}\,c^\dag_i\,c_j}\right|_{p\;\text{fermions}}&=\sum_p
  \lambda^p \sum_{i_1,...,i_p} \det_p \,U^c_{i_1,...,i_p}\\
&=\det\big(1+\lambda U^c(t)\big).
\end{aligned}
\end{equation}

Putting all together, the generating function of the 
contribution of a classical periodic orbit `$c$' is:
\begin{equation}
  \label{eqn:generatricetraceuneorbite}
  T^c(\lambda,t)=\frac{\det\big(1+\lambda
    U^c(t)\big)}{|\det\big(1-U^c(t)\big)|}, 
\end{equation}
and we hence have:
\begin{equation}
  \label{eqn:tracetotale}
  T(\lambda,t) = \sum_{\substack{\text{noiseless}\\\text{orbits
        c}}}\frac{\det\big(1+\lambda 
    U^c(t)\big)}{|\det\big(1-U^c(t)\big)|}.
\end{equation}

\section{}
\label{app:gaussiandev}

Let us consider a single saddle point (we drop the label `$c$'), which
we assume is at the origin. We develop $V$ to second order in $q_i$,
and may assume further that 
$\frac{\partial^2 V}{\partial q_i \partial q_j}(0)$ is
diagonalized. We can hence treat each mode separately, so for ease of
notation we drop the sub-indices $i,j$.  We wish to construct for each
mode of each saddle point, an eigenvector that is zero to this order:
\begin{equation}
  \label{eqn:equationgaussienne}
  H^{\text{c}}  \ket{{\psi_h}}=0,
\end{equation}
that is, for some $\lambda$:
\begin{equation}
  \left\{\begin{aligned}
  \left(- \gamma T \frac{\partial^2}{\partial
    p^2} -\frac{\gamma}{2} + \frac{\gamma}{4 T} p^2 - V'' q
  \frac{\partial}{\partial p}+p \frac{\partial }{\partial
    q}\right)  \ket{\psi_b^{h R}} &= \lambda  \ket{\psi_b^{h R}} \\
  \left(V'' b^\dag a+
  \gamma b^\dag b - 
  a^\dag b \right) \ket{{\psi_f^{h R}}} &= -\lambda \ket{{\psi_f^{h R}}}.
  \label{ccoo}
  \end{aligned}\right.
\end{equation}

Let us first have a look at the fermionic part. In the basis $\vidk,
\ket{a^\dag} , \ket{b^\dag}, \ket {a^\dag b^\dag}$, the fermionic part 
of (\ref{ccoo})  reads:
\begin{equation}
  H_{\text{ferm.}} = 
  {\tiny{\begin{pmatrix} 0 & 0 & 0 & 0\\
    0 & 0 & -1 & 0 \\ 0 & V'' & \gamma & 0\\ 0&0&0&\gamma
  \end{pmatrix}}}.
\end{equation}
The spectrum and  the corresponding eigenvectors are easily obtained:
\begin{equation}
  \begin{aligned}
  \lambda=0 &\leftrightarrow \ket{{\psi_f^{h R}}}=\vidk\\
\lambda=\frac{\gamma}{2}\pm\frac{1}{2}\sqrt{\gamma^2 - 4
    V''}&\leftrightarrow \ket{{\psi_f^{h R}}}=-\gamma \pm \sqrt{\gamma^2-4 V''})\ket{a^\dag}+2 V'' 
  \ket{b^\dag}\\ 
\lambda=\gamma &\leftrightarrow \ket{{\psi_f^{h R}}}=\ket{a^\dag
  b^\dag}.
\end{aligned}
\end{equation}

Regarding the bosonic part of (\ref{ccoo}), the Gaussian form of the
eigenvector allows 
us to compute the l.h.s:
\begin{equation}
  \begin{aligned}
    H_{\text{bos.}} \ket{\psi_b^{h R}} &= \left( (B_{p p}-\frac{1}{2}) \gamma  +
  (\frac{\gamma}{4} -\gamma B_{p p}^2 -B_{p q})
  \frac{p^2}{T} +(V'' B_{p q} - \gamma B_{p q}^2)
  \frac{q^2}{T}\right. \\&\quad\left.+ (-2 \gamma B_{p q} B_{p p} + V'' B_{p
    p} +B_{q q}) \frac{p q}{T}\right)\ket{\psi_b^{h R}}.
  \end{aligned}
\end{equation}
The prefactor of the quadratic terms must be equal to zero. On the
other hand, the Gaussian must be well normalized, i.e. the
eigenvectors of $B$ must be positive. After a tedious but
straightforward calculation, two solutions are possible, depending on
the sign of $V''$:
\begin{itemize}
\item if $V''>0$, then $B_{pp}=\frac{1}{2}$, $B_{pq}=0$ and
  $B_{q q}=\frac{V''}{2}$. This corresponds to an eigenvalue
  $0$ for the bosonic part, and the corresponding fermionic part is
  thus the vacuum:
\begin{equation}
  \ket{{\psi_{V''>0}^{h R}}} = e^{-\frac{1}{4 T} (p^2 + V'' q^2)}
  \otimes \vidk;
\end{equation}
\item if $V''<0$, then $B_{pp}=\frac{\sqrt{\gamma^2 -4 V''}}{2 \gamma
    }$, $B_{pq}=\frac{V''}{\gamma
    }$ and $B_{q q}=-V'' \frac{\sqrt{\gamma^2-4
    V''}}{2 \gamma }$. This corresponds to an
  eigenvalue $-\frac{\gamma}{2}+\frac{\sqrt{\gamma^2-4 V''}}{2}$
  for the bosonic part, which is compensated by the corresponding
  fermionic part $(-\gamma - \sqrt{\gamma^2-4 V''})\ket{a^\dag}+2 V''
  \ket{b^\dag}$:
  \begin{equation}
    \ket{{\psi_{V''<0}^{h R}}} = e^{-\frac{\sqrt{\gamma^2 -4 V''}}{4
	\gamma T} (p^2  - V'' q^2)-\frac{1}{2 \gamma T} V'' p q}
    \otimes \left((-\gamma - \sqrt{\gamma^2-4 V''})\ket{a^\dag}+2 V''
    \ket{b^\dag}\right).
  \end{equation}
\end{itemize}

The same development for $H^{ \dag}$ leads to:
\begin{itemize}
\item if $V''>0$, then $B_{pp}=\frac{1}{2}$, $B_{pq}=0$ and
  $B_{q q}=\frac{V''}{2}$. This corresponds to an eigenvalue
  $0$ for the bosonic part, and the corresponding fermionic part is
  thus the vacuum:
  \begin{equation}
    \ket{{\psi_{V''>0}^{h L}}} = e^{-\frac{1}{4 T} (p^2 + V'' q^2)}
  \otimes \vidk;
\end{equation}
\item if $V''<0$, then $B_{pp}=\frac{\sqrt{\gamma^2 -4 V''}}{2 \gamma
    }$, $B_{pq}=-\frac{V''}{\gamma
    }$ and $B_{q q}=-V'' \frac{\sqrt{\gamma^2-4
    V''}}{2 \gamma }$. This corresponds to an
  eigenvalue $-\frac{\gamma}{2}+\frac{\sqrt{\gamma^2-4 V''}}{2}$,
  for the bosonic part, which is compensated by the corresponding
  fermionic part $(\gamma + \sqrt{\gamma^2-4 V''})\ket{a^\dag}+2 
  \ket{b^\dag}$:
  \begin{equation}
    \ket{{\psi_{V''<0}^{h L}}} = e^{-\frac{\sqrt{\gamma^2 -4 V''}}{4
	\gamma T} (p^2  - V'' q^2)+\frac{1}{2 \gamma T} V'' p q}
    \otimes \left((\gamma + \sqrt{\gamma^2-4 V''})\ket{a^\dag}+2 
    \ket{b^\dag}\right).
  \end{equation}
\end{itemize}

The structure of the spectrum can be directly seen in the
eigenvectors: a stable direction corresponds to the fermionic vacuum
while an unstable direction corresponds to a
one-fermion state. We have also shown that both the right and left
eigenvectors with zero eigenvalue are Gaussians to this order in the
new basis.

\bibliographystyle{apsrev}
\bibliography{/home/tailleur/Latex/biblio.bib}
\end{document}